\documentclass[prd,tightenlines,nofootinbib,superscriptaddress]{revtex4}

\usepackage{amsfonts,amssymb,amsthm,bbm}

\usepackage{amsmath}

\usepackage{hyperref}

\usepackage{subcaption}
\usepackage{color,psfrag}
\usepackage[dvips]{graphicx}

\usepackage{tikz}
\usetikzlibrary{calc,patterns,angles,quotes}
\usetikzlibrary{decorations.pathmorphing}
\usetikzlibrary{shapes.geometric}
\usetikzlibrary{arrows,decorations.markings}

\newcommand{\C}{{\mathbb C}}
\newcommand{\N}{{\mathbb N}}
\newcommand{\R}{{\mathbb R}}

\newcommand{\cA}{{\mathcal A}}

\newcommand{\cG}{{\mathcal G}}

\newcommand{\cD}{{\mathcal D}}

\newcommand{\cS}{{\mathcal S}}

\newcommand{\cI}{{\mathcal I}}

\newcommand{\cZ}{{\mathcal Z}}

\newcommand{\SU}{\mathrm{SU}}

\newcommand{\be}{\begin{equation}}
\newcommand{\ee}{\end{equation}}
\newcommand{\beq}{\begin{eqnarray}}
\newcommand{\eeq}{\end{eqnarray}}
\newcommand{\bes}{\begin{eqnarray}}
\newcommand{\ees}{\end{eqnarray}}

\newcommand{\mat} [2] {\left ( \begin{array}{#1}#2\end{array} \right ) }
\newcommand{\sixj} [6] {\left\{ \begin{array}{ccc} {#1}&{#2}&{#3}\\{#4}&{#5}&{#6}\end{array} \right\} }

\newcommand{\bin} [2] {\left (\begin{array}{c}#2\\#1\end{array} \right ) }

\newcommand{\tr}{{\mathrm{Tr}}}
\newcommand{\f}{\frac}

\def\nn{\nonumber}
\def\pp{\partial}

\def\eps{\epsilon}

\def\rd{\mathrm{d}}

\def\tY{\widetilde{Y}}

\def\tGamma{\widetilde{\Gamma}}

\newcommand{\alink}[4]
{\draw[decoration={markings,mark=at position 0.6 with {\arrow[scale=1.5,>=stealth]{>}}},postaction={decorate}] (#1) -- node[#3,pos=.5]{$#4$}(#2)}
\newcommand{\link}[2]
{\draw[decoration={markings,mark=at position 0.6 with {\arrow[scale=1.5,>=stealth]{>}}},postaction={decorate}] (#1) --(#2)}
\newcommand{\ink}[2]
{\draw[dashed] (#1) --(#2)}

\def\centerarc[#1](#2)(#3:#4:#5)
    { \draw[#1] ($(#2)+({#5*cos(#3)},{#5*sin(#3)})$) arc (#3:#4:#5); }
    
\def\centerarcnodes[#1](#2)(#3:#4:#5)(#6,#7)
    {\coordinate(#6) at ($(#2)+({#5*cos(#3)},{#5*sin(#3)})$);
    \coordinate(#7) at ($(#2)+({#5*cos(#4)},{#5*sin(#4)})$);
    \draw[#1] ($(#2)+({#5*cos(#3)},{#5*sin(#3)})$) arc (#3:#4:#5); }

\def\angcircle(#1)(#2)(#3:#4) {\coordinate(#1) at ($(#2)+({#4*cos(#3)},{#4*sin(#3)})$); }

\begin{document}

\title{Self-duality of the 6j-symbol and Fisher zeros for the Tetrahedron}

\author{{\bf Valentin Bonzom}}\email{bonzom@lipn.univ-paris13.fr }
\affiliation{LIPN,  CNRS UMR   7030,  Institut  Galil\'ee,  Universit\'e  Paris  13,   93430  Villetaneuse,  France}

\author{{\bf Etera R. Livine}}\email{etera.livine@ens-lyon.fr}
\affiliation{Perimeter Institute for Theoretical Physics, 31 Caroline Street North, Waterloo, Ontario, Canada N2L 2Y5}
\affiliation{Universit\'e de Lyon, ENS de Lyon,  Laboratoire de Physique, CNRS UMR 5672, F-69342 Lyon, France}

\date{\today}

\begin{abstract}

The relation between the 2d Ising partition function and spin network evaluations, reflecting a bulk-boundary duality between the 2d Ising model and 3d quantum gravity, promises an exchange of results and methods between statistical physics and quantum geometry.
We apply this relation to the case of the tetrahedral graph. First, we find that the high/low temperature duality of the 2d Ising model translates into a new self-duality formula for Wigner's 6j-symbol from the theory of spin recoupling. Second, we focus on the duality between the large spin asymptotics of the 6j-symbol and Fisher zeros. Using the  Ponzano-Regge formula for the asymptotics for the 6j-symbol at large spins in terms of the tetrahedron geometry, we obtain a geometric formula for the zeros of the (inhomogeneous) Ising partition function in terms of triangle angles and dihedral angles in the tetrahedron. While it is well-known that the 2d intrinsic geometry can be used to parametrize the critical point of the Ising model, e.g. on isoradial graphs, it is the first time to our knowledge that the extrinsic geometry is found to also be relevant.
This outlines a method towards a more general geometric parametrization of the Fisher zeros for the 2d Ising model on arbitrary graphs.

\end{abstract}

\maketitle
\tableofcontents

\section*{Introduction}

Quantum gravity models, defined as sums over random geometries resulting from gluing elementary blocks of geometry, naturally have a strong interplay with statistical physics. For instance, 2d quantum gravity can be formulated as a sum over random 2d triangulations implemented through matrix models (see e.g. \cite{Kazakov:1986hu,DiFrancesco:2004qj}) and such matrix ensembles have revealed themselves to be general templates for large classes of statistical physics models \cite{doi:10.1137/1009001}. Beside the recent extension of this line of research to quantum gravity in higher dimensions and tensor models \cite{Bonzom:2011zz,Bonzom:2011ev,Rivasseau:2016zco,Witten:2016iux}, the interaction and bond between the fields of quantum gravity and statistical physics have also been renewed by the study of holographic dualities: the dynamics of observables and correlations of geometry or quantum geometry within a region of space-time is conjectured to be entirely by a theory living on the boundary of that region. The boundary theory is usually thought to be a conformal field theory, or suitable deformation, in a continuum approach (e.g. in the framework of AdS/CFT correspondence), or defined as discrete models from condensed matter and statistical physics in a discrete boundary geometry setting (e.g. in quasi-local approaches to gravity/gauge holographic duality).

\medskip

\begin{figure}[h!]
\includegraphics[scale=.5]{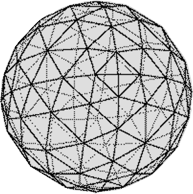}

\caption{\label{fig:3ball}
An example of triangulation of the 2-sphere as the boundary of a 3-ball. The Ponzano-Regge amplitude associated to a triangulation of the 3-ball does not depend on the details of the bulk triangulation but only depends on the boundary triangulation and on the boundary state \cite{PR1968,Freidel:2004vi,Freidel:2005bb,Barrett:2008wh}. The boundary state consists in the assignment of half-integers - or spins - to each edge of the boundary triangulation indicating their length quantized in Planck unit. The Ponzano-Regge amplitude is then given by the corresponding spin network evaluation. The generating function of those spin network evaluations has been shown to be related to the Ising model living on the planar graph dual to the boundary triangulation \cite{Bonzom:2015ova,Westbury,Freidel:2012ji}.}
\end{figure}

Such an example of bulk-boundary duality bridging between quantum gravity and statistical physics techniques is a duality formula between 3d quantum gravity, defined by the Ponzano-Regge model as a  path integral over discrete geometries \cite{PR1968,Freidel:2004vi,Freidel:2005bb,Barrett:2008wh}, and the (inhomogeneous) 2d Ising model, as shown in \cite{Bonzom:2015ova}. More precisely, it was shown, either through  direct computation or using a graded gauge symmetry, that the Ponzano-Regge amplitude on a region of space-time with a 3-ball topology is equal to the squared inverse of the 2d Ising model living on the boundary triangulation, as illustrated on fig.\ref{fig:3ball}.

This result promises the possibility of using known results and methods already developed for the Ising model in order to  study the  phase diagram of the discrete geometry in 3d quantum gravity and analyze potential phase transitions as sketched in  \cite{Bonzom:2015ova,Dittrich:2013jxa}. But it seems plausible to also go the other way round: apply techniques and results from quantum gravity to revisit the Ising model.
Indeed, as hinted by preliminary results in \cite{Bonzom:2015ova}, we could use the saddle point methods developed to investigate the semi-classical regime of spinfoams \cite{Barrett:1998gs,Livine:2007vk,Barrett:2010ex} and derive the asymptotics of spin network evaluations \cite{Dowdall:2009eg,costantino2015generating} to provide a geometrical formula for the zeros of the Ising model.

\medskip

Here we test this method on the tetrahedron graph and study the relation between the generating function for 6j-symbols, on the quantum gravity side, with the Ising model on the tetrahedron, on the statistical physics side. On one hand, we strive for a better understanding of the properties of the 6j-symbols from the well-studied structures of the Ising model partition function. On the other hand, we hope for an interpretation of the Ising partition function, correlations and zeros, in terms of 3d (quantum) geometry.

In particular, we present two original results:
\begin{itemize}

\item Introducing the generating function for the 6j-symbols of the recoupling theory of spins,
\be
\cZ_{T}[\{Y_e\}]
\,=\,
\sum_{\{j_{e\}}} \prod_{v}\Delta_v(j_e)\,
\sixj{j_{1}}{j_{2}}{j_{3}}{j_{4}}{j_{5}}{j_{6}} \,\prod_{e}Y_{e}^{2j_{e}}
\,,
\ee
where the spins $j_{e}\in\f\N2$ are half-integers, $\left\{\begin{smallmatrix} j_1 & j_2 & j_3\\ j_4 & j_5 & j_6 \end{smallmatrix}\right\}$ the associated 6j-symbol, and the combinatorial coefficients $\Delta_v(a,b,c)$ are defined explicitly in section \ref{sec:6jdef}, we translate the high/low temperature duality of the 2d Ising model into a self-duality formula:
\be
\prod_{e=1}^6 (1+Y_{e})^2\,\cZ_{T}[\{Y_e\}]=2^6\,\cZ_{T^*}[\{Y_e^*\}]\,
\ee
where we relate the generating function for the tetrahedron $T$ to its dual $T^*$ where opposite edges have been exchanged, $(1,2,3)\leftrightarrow (4,5,6)$, and the variables $Y_{e}$ are sent to their dual version,
\be
Y_{e}^*=\f{1-Y_{e}}{1+Y_{e}}
\,,\qquad
Y_{e}=\f{1-Y_{e}^*}{1+Y_{e}^*}
\,.
\ee
This can also be written directly as a self-duality identity for the 6j-symbols,
\be
2^6 \sum_{\{j_e\}} \begin{Bmatrix} j_1& j_2& j_3\\ j_4& j_5& j_6\end{Bmatrix} \prod_v \Delta_v(j_e) \prod_e (-1)^{2k_{e}} T(2j_e+1, 2k_e+1) = \begin{Bmatrix} k_4& k_5& k_6\\ k_1& k_2& k_3\end{Bmatrix} \prod_{v^*} \Delta_{v^*}(k_e).
\,,
\ee
where the $(T(2j+1, 2k+1))_{k\in\mathbbm{N}/2}$ are figurate numbers for the $(2j+1)$-dimensional cross polytopes (reference \href{http://www.oeis.org/A142978}{A142978 on oeis.org}). The formula above is rather similar to the self-duality formula written for the $q$-deformed 6j-symbols \cite{Freidel:2006qv}, but the latter involves the (discrete) Fourier transform of the {squared} 6j-symbol while the new duality formula applies directly to the 6j-symbol.

\item Using the asymptotic formula for the 6j-symbols at large spins, we study the stationary points of the series defining the generating function $\cZ_{T}[\{Y_e\}]$, which gives us a formula for a real section of the sets of Fisher zeros for the Ising partition function $\cI_{T}[\{Y_e\}]$ on the tetrahedron graph $T$ in terms of tetrahedron geometry:
\be
Y_{e}^c
=
e^{\eps \f{\theta_{e}}2}
\,
\sqrt{\tan\f{\phi_1^e}2 \tan\f{\phi_2^e}2}
\,,\qquad
\cI_{T}[\{Y_e^c\}]=0
\,,
\ee
where $\eps$ is an overall sign, $\theta_{e}$ is the dihedral angle between the two triangles sharing the edge $e$ and the $\phi_{1,2}^e$ are the angles in those two triangles opposite to the edge $e$.
This reduces to the well-established formula for critical Ising couplings for isoradial graphs \cite{BoutillierDeTiliereSurvey},
$Y_{e}^c=\tan{\phi_{e}}/2$, if we consider a isoradial tetrahedron (the circumscribed circles of the triangles have all equal radii) and  ignore the phase coming from the dihedral angles. So the novelty of this formula especially resides in the phase term and the dihedral angles reflecting the embedding of the 2d graph into the 3d space, i.e. its extrinsic geometry.

Moreover we show how the continuation of this formula to complex edge lengths provides a parametrization of all Fisher zeros, i.e. all complex solutions to $\cI_{T}[\{Y_e^c\}]=0$. This provides a geometric formula for the critical couplings of the Ising model.
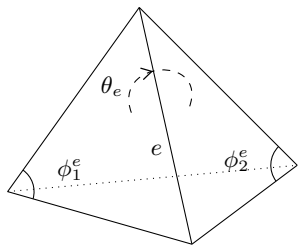
\begin{figure}[h!]

\centering

\begin{tikzpicture}[scale=0.7]

\coordinate(A) at (0,0);
\coordinate(B) at (2.5,3.5);
\coordinate(C) at (3.5,-1);
\coordinate(D) at (5.5,.5);

\draw (A)--(B) ;
\draw (A)--(C) ;
\draw (D)--(B) ;
\draw (D)--(C) ;
\draw[dotted] (A)--(D) ;

\pic [draw, "$\phi_1^e$", angle eccentricity=2.5,angle radius=10] {angle = C--A--B};
\pic [draw, "$\phi_2^e$", angle eccentricity=2.3,angle radius=10] {angle = B--D--C};

\draw (B)-- node[pos=0.6,left]{$e$} (C);

\draw[dashed,in=+70,out=+110,looseness=2.5,decoration={markings,mark=at position 0.45 with {\arrow[scale=1.2,>=angle 60]{>}}},postaction={decorate}] (2.35,1.5) to node[left,pos=.2]{$\theta_{e}$} (3.45,1.6);

\end{tikzpicture}

\caption{
Geometry of a tetrahedron: 2d opposite angles and 3d dihedral angle associated to the edge $e$.
} 
\label{fig:Tgeometry}

\end{figure}

\end{itemize}
We consider the present work as a first check of this method before applying it to derive a more general geometric formula for the Fisher zeros of the Ising partition function on arbitrary planar graphs from the asymptotics of spin network evaluations studied in 3d quantum gravity. We also hope,  in the future, to apply the high/low temperature duality of the Ising model to obtain UV/IR duality formulas for the quantum gravity amplitudes and maybe help better characterize critical regimes of 3d quantum geometry.

\section{Duality between Spin Networks and 2d Ising}

\subsection{Spin network evaluation, loop polynomial and Ising partition function}

The duality between 3D quantum gravity\footnotemark{} and the 2D Ising model, described in \cite{Bonzom:2015ova}, is expressed as a relation between the 3D quantum gravity partition function on the 3-ball with a non-trivial boundary state on the 2D sphere and the 2D Ising partition function on the same boundary. 
\footnotetext{We are considering 3D Euclidean quantum gravity, meaning that the space-time metric has a positive signature (+++) at the classical level. We are not Wick-rotating the path integral and are considering the quantum mechanical path integral summing over all 3D metrics with positive signature, $\int \rd {}^{3d}g\,\exp[i\,S_{grav}[ {}^{3d}g]]$. At the quantum level, the Ponzano-Regge discrete path integral will nevertheless contain exponentially-suppressed  contributions from Lorentzian metric with non-definite signature (-++) \cite{Barrett:1993db}.}

\medskip

On the one hand, 3D quantum gravity is a topological field theory, whose path integral is described at the discrete level by the Ponzano-Regge state-sum model. The Ponzano-Regge model defines amplitudes for the geometry of every 3D triangulation. Working on the 3-ball with a 2-sphere boundary, the topological invariance of the Ponzano-Regge model can be used to show that the amplitude does not depend on the details of the 3D triangulation in the 3-ball's bulk but does actually depend non-trivially on the 2D boundary triangulation (see e.g. \cite{Freidel:2004vi,Freidel:2004nb,Freidel:2005bb}). The boundary state is defined as a spin network and the Ponzano-Regge formula for the 3D quantum gravity amplitude with boundary is given by what is called the evaluation of the spin network.
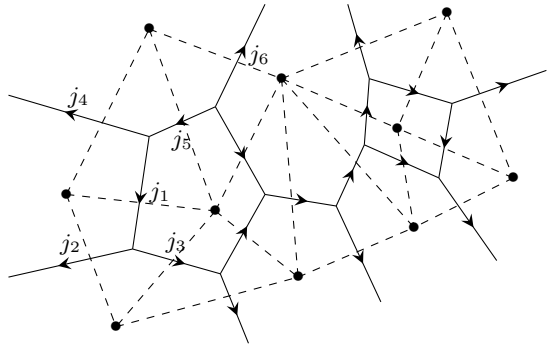
\begin{figure}[h!]

\begin{tikzpicture}[scale=2.2]

\coordinate(a) at (0,0) ;
\coordinate(b) at (.5,1);
\coordinate(c) at (.9,-.1);
\coordinate(d) at (.3,-.8);
\coordinate(e) at (1.3,.7);
\coordinate(f) at (2.3,1.1);
\coordinate(g) at (2,.4);
\coordinate(h) at (2.7,.1);
\coordinate(i) at (2.1,-.2);
\coordinate(j) at (1.4,.-.5);

\coordinate(abc) at (.5,0.35) ;
\coordinate(acd) at (.4,-0.33);
\coordinate(bce) at (.9,0.53);
\coordinate(cdj) at (.93,-0.48);
\coordinate(cej) at (1.2,0);
\coordinate(eij) at (1.63,-0.07);
\coordinate(eig) at (1.8,0.3);
\coordinate(egf) at (1.83,0.7);
\coordinate(fgh) at (2.33,0.55);
\coordinate(igh) at (2.25,0.1);

\coordinate(ab) at (-0.35,0.6);
\alink{abc}{ab}{above}{j_{4}};
\coordinate(be) at (1.2,1.15);
\alink{bce}{be}{right}{j_{6}};
\coordinate(ad) at (-0.35,-0.5);
\alink{acd}{ad}{above}{j_{2}};
\coordinate(dj) at (1.1,-0.9);
\link{cdj}{dj};
\coordinate(ij) at (1.9,-0.65);
\link{eij}{ij};
\coordinate(ih) at (2.6,-0.4);
\link{igh}{ih};
\coordinate(fh) at (2.9,0.75);
\link{fgh}{fh};
\coordinate(ef) at (1.7,1.15);
\link{egf}{ef};

\alink{abc}{acd}{right}{j_{1}};
\alink{bce}{abc}{below}{j_{5}};
\alink{acd}{cdj}{above}{j_{3}};
\link{cdj}{cej};
\link{bce}{cej};
\link{cej}{eij};
\link{eij}{eig};
\link{eig}{egf};
\link{egf}{fgh};
\link{fgh}{igh};
\link{eig}{igh};

\draw (a) node {$\bullet$} ;
\draw (b) node {$\bullet$};
\draw (c) node {$\bullet$};
\draw (d) node {$\bullet$};
\draw (e) node {$\bullet$};
\draw (f) node {$\bullet$};
\draw (g) node {$\bullet$};
\draw (h) node {$\bullet$};
\draw (i) node {$\bullet$};
\draw (j) node {$\bullet$};


\ink{a}{b};
\ink{a}{c};
\ink{a}{d};
\ink{b}{c};
\ink{c}{d};
\ink{b}{e};
\ink{e}{f};
\ink{c}{e};
\ink{e}{g};
\ink{f}{g};
\ink{f}{h};
\ink{h}{g};
\ink{h}{i};
\ink{g}{i};
\ink{j}{i};
\ink{d}{j};
\ink{c}{j};
\ink{e}{j};
\ink{e}{i};

				
\end{tikzpicture}

\caption{Connected region of a planar, 3-valent, oriented graph $\Gamma$ and its dual triangulation $\Gamma^*$ drawn in dotted lines. We choose a Kasteleyn orientation on $\Gamma$, so that there is an odd number of edges oriented clockwise around each face of the graph. A spin network on $\Gamma$ consists in the assignment of spins $j_{e}\in\N/2$ to each link $e$ of the graph. They are interpreted as the length (in Planck unit) of the dual edge $e^*$ in the dual triangulation $\Gamma^*$.}
\label{fig:spinnetwork}
\end{figure}

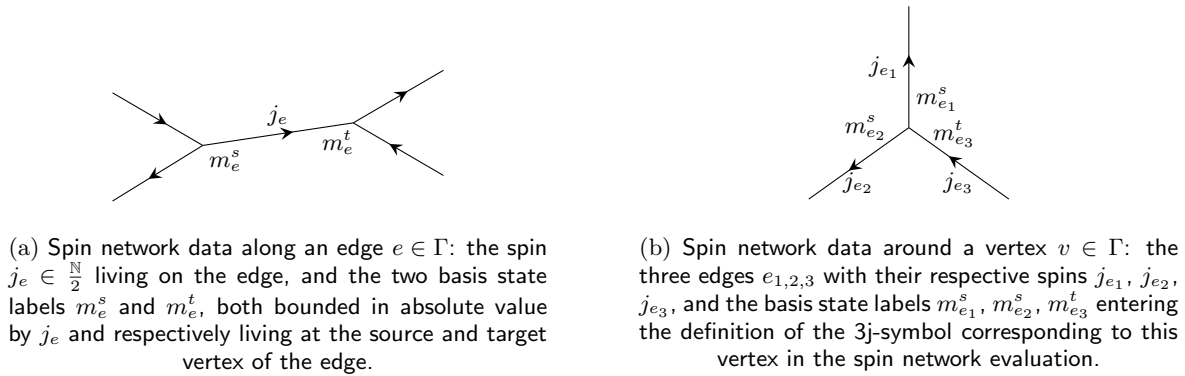
\begin{figure}[h!]

\begin{subfigure}[t]{.4\linewidth}
\centering
\begin{tikzpicture}[scale=2]

\coordinate(a) at (0,0);
\coordinate(b) at (1,0.15);

\alink{a}{b}{above}{j_{e}};
\draw (a)++(0.15,-0.1)node {$m_{e}^s$} ;
\draw (b)++(-0.1,-0.13) node {$m_{e}^t$} ;

\coordinate(a1) at (-0.6,0.35);
\coordinate(a2) at (-0.6,-0.37);
\coordinate(b1) at (1.6,0.5);
\coordinate(b2) at (1.6,-0.2);

\link{a1}{a};
\link{a}{a2};
\link{b2}{b};
\link{b}{b1};

\end{tikzpicture}

\caption{Spin network data along an edge $e\in\Gamma$: the spin $j_{e}\in\f\N2$ living on the edge, and the two basis state labels $m_{e}^s$ and $m_{e}^t$, both bounded in absolute value by $j_{e}$ and respectively living at the source and target vertex of the edge.}

\end{subfigure}
\hspace*{10mm}
\begin{subfigure}[t]{.4\linewidth}
\begin{tikzpicture}[scale=1.9]

\coordinate(o) at (0,0);
\coordinate(a) at (0,0.85);
\coordinate(b) at (-0.7,-0.5);
\coordinate(c) at (0.7,-0.5);

\alink{o}{a}{left}{j_{e_{1}}};
\alink{o}{b}{below}{j_{e_{2}}};
\alink{c}{o}{below}{j_{e_{3}}};

\draw (o)++(0.2,0.2) node{$m_{e_{1}}^s$};
\draw (o)++(-0.3,0) node{$m_{e_{2}}^s$};
\draw (o)++(0.32,-0.04) node{$m_{e_{3}}^t$};

\end{tikzpicture}

\caption{Spin network data around a vertex $v\in\Gamma$: the three edges $e_{1,2,3}$ with their respective spins $j_{e_{1}}$, $j_{e_{2}}$, $j_{e_{3}}$, and the basis state labels $m_{e_{1}}^s$, $m_{e_{2}}^s$, $m_{e_{3}}^t$ entering the definition of the 3j-symbol corresponding to this vertex in the spin network evaluation.
}

\end{subfigure}

\caption{The basic building blocks of a spin network evaluation on the edges and vertices on a graph $\Gamma$.
\label{fig:spinevaluation}}

\end{figure}

So let us consider a planar, 3-valent, oriented graph $\Gamma$.
We dress each edge $e\in\Gamma$ with a spin $j_{e}\in\f\N2$, which corresponds to an irreducible unitary representation of the Lie group $\SU(2)$.  Geometrically, the planar 3-valent graph $\Gamma$ defines a dual 2D planar triangulation $\Gamma^*$ and the spins are interpreted as the edge lengths of $\Gamma^*$ quantized in Planck length units, $l_{e}=j_{e}\ell_{Planck}$, as illustrated on fig.\ref{fig:spinnetwork}. To each vertex $v\in\Gamma$, we attach the 3j-symbol, from the spin recoupling theory, between the spins $j_{e_{1}^v}$, $j_{e_{2}^v}$, $ j_{e_{3}^v}$, living on the edges attached to the vertex $v$. Then the spin network evaluation on the oriented graph is defined as the contraction of all the 3j-symbols (see \cite{Bonzom:2015ova} for more details), as illustrated on fig.\ref{fig:spinevaluation}:
\be
s_{\Gamma}\big[\{j_{e}\}_{e\in\Gamma}\big]
=
\sum_{\{m_{e}\}}
\prod_{e}(-1)^{j_{e}-m_{e}}\prod_{v}
\mat{ccc}{j_{e_{1}^v} &j_{e_{2}^v} &j_{e_{3}^v} \\ \eps_{1}^v m_{e_{1}^v}&\eps_{2}^v m_{e_{2}^v}&\eps_{3}^v m_{e_{3}^v}}
\,,
\ee
where the sign $\eps_{a}^v$, with $a=1,2,3$, records the orientation of the edge $e_{a}^v$ with respect to the vertex $v$, i.e. $\eps_{a}^v=+1$ if the edge is incoming ($v=t(e)$) and $\eps_{a}^v=-1$ if the edge is outgoing ($v=s(e)$).
We then define the generating function for these spin network evaluations on the graph $\Gamma$ by introducing edge couplings $Y_{e}$ dual to the spins and summing over the spins $j_{e}$:
\be
\cZ_{\Gamma}[\{Y_{e}\}]
=
\sum_{\{j_{e}\}}
s_{\Gamma}[\{j_{e}\}]
\,\prod_{v}\sqrt{\f{(J_{v}+1)!}{\prod_{e\ni v}(J_{v}-2j_{e})!}}
\,\prod_{e}Y_{e}^{2j_{e}}
\,.
\ee
This can also be interpreted as the evaluation of coherent spin network states, defining coherent superpositions of spins on the 2D boundary \cite{Freidel:2012ji,Bonzom:2012bn,Bonzom:2015ova}.

\medskip

On the other hand, we consider the 2D Ising model defined on the same graph $\Gamma$, with ``spins'' $\sigma_{v}=\pm$ living at every vertex $v\in\Gamma$ with nearest neighbor interaction along the graph $\Gamma$ with edge couplings $y_{e}$. The Ising partition function reads:
\be
\cI_{\Gamma}[\{y_{e}\}]
=
\sum_{\sigma_{v}=\pm} e^{\sum_{e}y_{e}\sigma_{s(e)}\sigma_{t(e)}}
=
\left(
\prod_{e}
\cosh y_{e}
\right)
\,
\sum_{\sigma_{v}=\pm}
\prod_{e}(1+\sigma_{s(e)}\sigma_{t(e)}\tanh y_{e})
\,.
\ee
Since the sum over odd products of the spins $\sigma_{v}$  vanishes, the edge factors $\sigma_{s(e)}\sigma_{t(e)}$ must necessarily warp around cycles of the graph in order to get products of $\sigma_{v}^2$ and obtain non-vanishing contributions. This gives the (high temperature) loop expansion of the Ising model:
\be
\cI_{\Gamma}[\{y_{e}\}]
=
2^V\left(
\prod_{e}
\cosh y_{e}
\right)
\,
P_{\Gamma}[\{\tanh y_{e}\}]
\,,\qquad\textrm{with}\quad
P_{\Gamma}[\{Y_{e}\}]
=
\sum_{\cG\subset \Gamma}\prod_{e\in\cG}Y_{e}
\,.
\ee
where $\cG$ runs over all even subgraphs of $\Gamma$, i.e. such that the valency of every vertex of $\cG$ is even. Since the original graph $\Gamma$ is 3-valent, the valency of vertices of $\cG$ can only be 0 or 2, meaning that $\cG$ is necessarily a union of disjoint loops on $\Gamma$, hence we call $P_{\Gamma}$ the  {\bf loop polynomial}.
The Ising partition function is given up to a pre-factor by an evaluation of the loop polynomial, $P_{\Gamma}[\{\tanh y_{e}\}]$.

\medskip

The duality formula between the spin network evaluation and the Ising model, proven in \cite{Westbury,Freidel:2012ji} and by the authors using a graded Lie algebra symmetry \cite{Bonzom:2015ova}, is\footnotemark{}:
\be
\label{westbury}
\cZ_{\Gamma}[\{Y_{e}\}]
=
\f1
{P_{\Gamma}[\{Y_{e}\}]^2}
\,,\qquad
\cZ_{\Gamma}[\{Y_{e}\}]\cI_{\Gamma}[\{y_{e}\}]^2
=
2^{2V}
\prod_{e}
\cosh^2 y_{e}
\,,
\qquad\textrm{with}\quad
Y_{e}=\tanh y_{e}
\,.
\ee
\footnotetext{
As shown in \cite{Bonzom:2015ova},the equality between the spin network generating function and the squared inverse of the Ising partition function holds if we choose a Kasteleyn orientation for the edges of the graph $\Gamma$ to define the spin network evaluation. Indeed, if we switch the orientation of an edge $e$, then the spin network evaluation $s_{\Gamma}[\{j_{e}\}]$ acquires a sign $(-1)^{2j_{e}}$, so that one should also switch the sign of the edge coupling $Y_{e}\rightarrow -Y_{e}$ in the spin network generating function $\cZ_{\Gamma}[\{Y_{e}\}]$.
\\
To fix this sign ambiguity, we choose a Kasteleyn orientation, which defines an orientation of the edges compatible with the planarity of the graph. Since the graph is assumed to be planar, we have a (counter clockwise) orientation for every face of the graph $\Gamma$. A Kasteleyn orientation of the edges is such that each face has an odd number of edges whose orientation does not match the one of the face.
}
Since $1/\cosh^2 y=1-\tanh^2y=1-Y^2$, we can rewrite this duality formula as:
\be
\prod_{e}(1-Y_{e}^2)
\,\cZ_{\Gamma}[\{Y_{e}\}]\,\cI_{\Gamma}[\{y_{e}\}]^2
=
2^{2V}\,,
\ee
or, in words, the generating function for spin network evaluation --the 3D quantum gravity amplitude with boundary-- is the squared inverse of the 2D Ising model.

\subsection{Low/high temperature expansion of the Ising model and duality}

A powerful technique to study the 2D Ising model is the duality between the low and high temperature regimes, which allows a direct derivation for the critical couplings for regular lattices, see e.g. \cite{Baxter:1982zz}. Indeed, while the Ising partition function admits a loop expansion in the high temperature regime as shown above, it also admits a loop expansion in the low temperature regime but on the dual graph $\Gamma^*$.

\begin{figure}[h!]

\begin{subfigure}[t]{.47\linewidth}
\centering
\begin{tikzpicture}[scale=1.8]

\coordinate(a) at (0,0) ;
\coordinate(b) at (.5,1);
\coordinate(c) at (.9,-.1);
\coordinate(d) at (.3,-.8);
\coordinate(e) at (1.3,.7);
\coordinate(f) at (2.3,1.1);
\coordinate(g) at (2,.4);
\coordinate(h) at (2.7,.1);
\coordinate(i) at (2.1,-.2);
\coordinate(j) at (1.4,.-.5);

\coordinate(abc) at (.5,0.35) ;
\coordinate(acd) at (.4,-0.33);
\coordinate(bce) at (.9,0.53);
\coordinate(cdj) at (.93,-0.48);
\coordinate(cej) at (1.2,0);
\coordinate(eij) at (1.63,-0.07);
\coordinate(eig) at (1.8,0.3);
\coordinate(egf) at (1.83,0.7);
\coordinate(fgh) at (2.33,0.55);
\coordinate(igh) at (2.25,0.1);

\coordinate(ab) at (-0.35,0.6);
\link{abc}{ab};
\coordinate(be) at (1.2,1.15);
\link{bce}{be};
\coordinate(ad) at (-0.35,-0.5);
\link{acd}{ad};
\coordinate(dj) at (1.1,-0.9);
\link{cdj}{dj};
\coordinate(ij) at (1.9,-0.65);
\link{eij}{ij};
\coordinate(ih) at (2.6,-0.4);
\link{igh}{ih};
\coordinate(fh) at (2.9,0.75);
\link{fgh}{fh};
\coordinate(ef) at (1.7,1.15);
\link{egf}{ef};

\link{abc}{acd};
\link{bce}{abc};
\link{acd}{cdj};
\link{cdj}{cej};
\link{bce}{cej};
\link{cej}{eij};
\link{eij}{eig};
\link{eig}{egf};
\link{egf}{fgh};
\link{fgh}{igh};
\link{eig}{igh};

\draw (a) node {$\bullet$} ;
\draw (b) node {$\bullet$};
\draw (c) node {$\bullet$};
\draw (d) node {$\bullet$};
\draw (e) node {$\bullet$};
\draw (f) node {$\bullet$};
\draw (g) node {$\bullet$};
\draw (h) node {$\bullet$};
\draw (i) node {$\bullet$};
\draw (j) node {$\bullet$};

\ink{a}{b};
\ink{a}{c};
\ink{a}{d};
\ink{b}{c};
\ink{c}{d};
\ink{b}{e};
\ink{e}{f};
\ink{c}{e};
\ink{e}{g};
\ink{f}{g};
\ink{f}{h};
\ink{h}{g};
\ink{h}{i};
\ink{g}{i};
\ink{j}{i};
\ink{d}{j};
\ink{c}{j};
\ink{e}{j};
\ink{e}{i};

\end{tikzpicture}

\caption{
Dual of a 3-valent graph. Even though the original graph $\Gamma$ is 3-valent, the dual graph nodes do not have any particular valency. The dual graph $\Gamma^*$ defines in this case a triangulation.
}

\end{subfigure}
\hspace*{10mm}
\begin{subfigure}[t]{.38\linewidth}
\begin{tikzpicture}[scale=1]

\foreach \i in {0,...,3}{
\foreach \j in {0,...,3}{
\draw[decoration={markings,mark=at position .9 with {\arrow[scale=1.5,>=stealth]{>}}},postaction={decorate}] (\i-.5,\j)--(\i+.5,\j);
\draw[decoration={markings,mark=at position .9 with {\arrow[scale=1.5,>=stealth]{>}}},postaction={decorate}] (\i,\j-.5)--(\i,\j+.5);
}
}

\foreach \i in {1,...,3}{
\draw[dashed] (-.5,\i-0.5) to (3.5,\i-0.5);
\draw[dashed] (\i-0.5,-.5) to (\i-0.5,3.5);
}

\foreach \i in {1,...,3}{
\foreach \j in {1,...,3}{
\draw (\i-0.5,\j-0.5) node {$\bullet$} ;
}
}

\end{tikzpicture}

\caption{The regular square lattice is self-dual in the sense that its dual graph   is once again a regular square lattice.
}

\end{subfigure}

\caption{
Examples of dual graphs: the initial oriented graph $\Gamma$ is drawn in plain lines, while the dual graph $\Gamma^*$ is drawn in dotted lines: each dual vertex $v^*$ corresponds to a face of the planar cellular complex defined by $\Gamma$, each dual edge $e^*$ corresponds to an edge of the original graph and links two neighboring faces of $\Gamma$.
\label{fig:dualgraph}}

\end{figure}
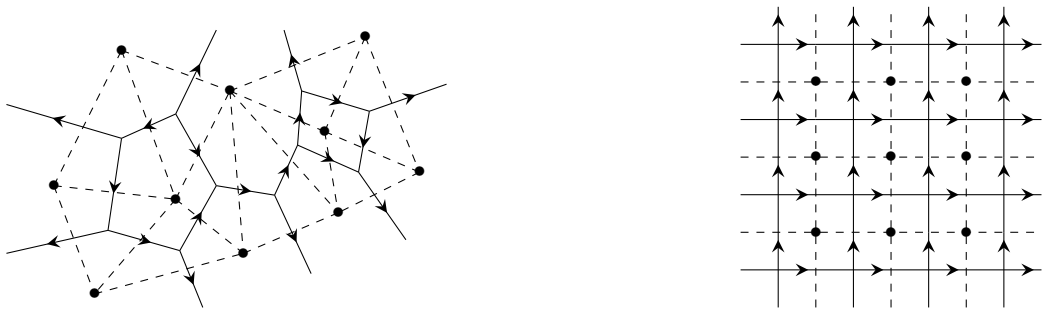

Indeed, the low temperature expansion of the Ising partition function is given by a cluster expansion: we describe each configuration of spins $\sigma_{v}$ by the connected clusters of positive spins and the connected clusters of negative spins. Each connected cluster of positive spin is bounded by a loop in the dual graph $\Gamma^*$. So we can restructure the Ising partition function as a sum over cycles in $\Gamma^*$.
More precisely, for a planar embedding of the graph $\Gamma$, we  define the dual graph $\Gamma^*$ by defining a dual vertex for each face and a dual edge transverse to an edge and thus linking a dual vertex to a neighboring one, as illustrated on fig.\ref{fig:dualgraph}. Then we write  $\cI_{\Gamma}[\{y_{e}\}]$ as a sum over loop systems on the dual graph $\Gamma^*$. Loops can meet at points, but not cross each, thus corresponding to an even subgraph of the dual graph $\Gamma^*$:
\be
\cI_{\Gamma}[\{y_{e}\}]
=
2
\prod_{e}
e^{y_{e}}
\,
\sum_{\cG^*\subset\Gamma^*}\prod_{e^*\in\cG^*}e^{-2y_{e}}
=
2
\prod_{e}
e^{y_{e}}
\,
P_{\Gamma^*}[\{Y^*_{e}\}]
\qquad\textrm{with}\quad
Y^*_{e}=e^{-2y_{e}}
\,,
\ee
where the pre-factor 2 comes from a global switch of signs of the spins (i.e. when loops in dual graph $\Gamma^*$ define border of clusters of  negative spins instead of clusters of positive spins).

The duality mapping $\cD$ between the edge couplings of the low and high temperature expansions is an involution:
\be
Y=\cD(Y^*)=\f{(1-Y^*)}{(1+Y^*)},
\qquad
Y^*=\cD(Y)=\f{(1-Y)}{(1+Y)}\,.
\ee
The fixed points of this map are $Y_{c}=-(1\pm\sqrt{2})$. These are the critical couplings for the 2d homogeneous Ising model on a square lattice (since the square lattice is its own dual graph).

This translates into the low/high temperature duality for the loop polynomials on the graph $\Gamma$ and its dual $\Gamma^*$:
\be
2^V\prod_{e}\cosh y_{e} P_{\Gamma}[\{\tanh y_{e}\}]
=
2
\prod_{e}e^{y_{e}}
P_{\Gamma^*}[\{e^{-2y_{e}}\}]
\,,
\ee
\be
\label{dualityonP}
2^{V-E-1}
\prod_{e}(1+Y^*_{e})\, P_{\Gamma}[\{Y_{e}\}]
=
P_{\Gamma^*}[\{Y^*_{e}\}]
\,,\qquad
P_{\Gamma}[\{Y_{e}\}]
=
2^{1-V}\prod_{e}(1+Y_{e})P_{\Gamma^*}[\{Y^*_{e}\}]
\,,
\ee
which holds since $(1+Y^*)(1+Y)=2$ by the duality map. These two formulas are  dual of each other, since ${(1-V)}$ plays the role of  $(V-E-1)$ for the dual graph:
\be
V^*-E^*-1
=
F-E-1
=
1-V
\,,
\nn
\ee
since we have assumed the planarity of the graphs and the Euler characteristic is fixed to $\chi_{\textrm{Euler}}=F-E+V=2$.

This promises a totally new and original relation between spin network evaluations on $\Gamma$ and spin network evaluations on the dual graph $\Gamma^*$, which would be especially interesting in the context of discretizations of 3D gravity combining both the triangulation and its topological dual as in \cite{Dupuis:2017otn,Delcamp:2018sef,Freidel:2018pbr}.
However this would require extending the duality formula \eqref{westbury} between the Ponzano-Regge amplitudes and the 2D Ising partition functions to graphs with node valency higher than 3.
Although this should definitely be studied at some point, this is not the purpose of the present work. Instead we will focus the tetrahedral graph, which is actually self-dual, i.e. the dual graph is once again a tetrahedral graph. This leads to a self-duality formula for the 6j-symbols and their generating function, as we will see in \ref{sec:tetra}.

\section{The $\Theta$ Graph and Dual Triangle Geometry}
\label{sec:triangle}

\begin{figure}[h!]

\begin{subfigure}[t]{.4\linewidth}
\centering
\begin{tikzpicture}[scale=1.3]

\draw[in=+90,out=+90,looseness=1,decoration={markings,mark=at position 0.6 with {\arrow[scale=1.5,>=stealth]{>}}},postaction={decorate}] (0,0) to node[above,pos=.5]{$j_{1}$} (1.5,0);
\draw[in=-90,out=-90,looseness=1,decoration={markings,mark=at position 0.6 with {\arrow[scale=1.5,>=stealth]{>}}},postaction={decorate}] (0,0) to node[above,pos=.5]{$j_{3}$}(1.5,0);
\alink{0,0}{1.5,0}{above}{j_{2}};

\draw (0,0) node {$\bullet$} ;
\draw (1.5,0) node {$\bullet$} ;

\end{tikzpicture}

\caption{
The $\Theta$-graph has two vertices linked by three edges. A spin network on $\Theta$ is the assignment of the spin $j_{i}$  to each of the three edges. The spin network evaluation consists in considering the two 3j-symbols corresponding to the two vertices and contracting them.}

\end{subfigure}
\hspace*{10mm}
\begin{subfigure}[t]{.45\linewidth}
\begin{tikzpicture}[scale=1]

\coordinate(a) at (0,0.85);
\coordinate(b) at (-0.8,-0.5);
\coordinate(c) at (0.8,-0.5);
\draw (a) node {$\bullet$} ;
\draw (b) node {$\bullet$} ;
\draw (c) node {$\bullet$} ;

\draw[in=80,out=-10,looseness=1,decoration={markings,mark=at position 0.6 with {\arrow[scale=1.5,>=stealth]{>}}},postaction={decorate}] (a) to node[above,pos=.45]{$j$}(c);
\draw[in=-20,out=200,looseness=1,decoration={markings,mark=at position 0.6 with {\arrow[scale=1.5,>=stealth]{>}}},postaction={decorate}] (c) to node[above,pos=.5]{$j$}(b);
\draw[out=100,in=190,looseness=1,decoration={markings,mark=at position 0.6 with {\arrow[scale=1.5,>=stealth]{>}}},postaction={decorate}] (b) to node[left,pos=.45]{$j$}(a);

\end{tikzpicture}

\caption{
The triangle graph $\Delta$ is made of three bivalent vertices linked by three edges forming a loop. A spin network on $\Delta$ assigns a spin on each of the three edges. Since the vertices are bivalent, those spins are necessarily all equal. The spin network evaluation is then simply the dimension of the corresponding representation, $\dim j=(2j+1)$.
}

\end{subfigure}

\caption{
The $\Theta$-graph and its dual graph, the triangle graph $\Delta$.
\label{fig:Theta-triangle}}

\end{figure}
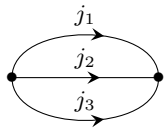
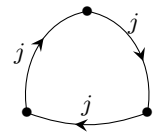

In this section, we start with the $\Theta$ graph and its dual three-edge loop (or  triangular graph), as illustrated on fig.\ref{fig:Theta-triangle}, to illustrate the duality between spin networks and the Ising model in a simple case as an appetizer for the more complete case of the tetrahedral case which we will analyze in the next section \ref{sec:tetra}. Here we will see that the spin network evaluations and the Ising partition function are both encoded in the triangle geometry.

\subsection{Generating function of spin network evaluations on the $\Theta$ graph}

We start with the simplest 3-valent graph, the $\Theta$ graph, made of two vertices linked by three edges as drawn on fig.\ref{fig:Theta-triangle}. A spin network is defined in terms of three spins $j_{1},j_{2},j_{3}$ living respectively on the $\Theta$'s three edges. Then the generating function is defined as a function of three edges couplings $Y_{1,2,3}$ weighting each spin:
\be
\cZ_{\Theta}[Y_{1,2,3}]
=
\sum_{j_{1,2,3}\in\f\N2}
\delta_{j_{1},j_{2},j_{3}}\,\f{(-1)^{j_{1}+j_{2}+j_{3}}(j_{1}+j_{2}+j_{3}+1)!}{(j_{1}+j_{2}-j_{3})!(j_{1}+j_{3}-j_{2})!(j_{2}+j_{3}-j_{1})!}
\prod_{i=1}^3Y_{i}^{2j_{i}}
\,.
\ee
The combinatorial factor $\delta_{j_{1},j_{2},j_{3}}$ can be defined as a integral over $\SU(2)$ of the products of the characters\footnotemark{} of the three $\SU(2)$ representations,
\be
\delta_{j_{1},j_{2},j_{3}}
=
\int_{\SU(2)}
\rd g\,
\chi_{j_{1}}(g)\chi_{j_{2}}(g)\chi_{j_{3}}(g)
\,,
\ee
and gives 1 if the three spins satisfy the triangular inequalities (and parity constraint) and 0 otherwise.
\footnotetext{The character in the spin $j$ is the function  $\chi_{j}$ of group elements $g\in\SU(2)$  giving  the trace of the Wigner-matrix representing the group element. It is a central function over $\SU(2)$, i.e. invariant under conjugation, and depends only on the conjugation class of $g$ labeled by its rotation angle:
\be
\chi_{j}(g)=\tr D^j(g)
\,,\qquad
\chi_{j}(e^{i\theta\sigma_{3}})=\f{\sin(2j+1)\theta}{\sin\theta}=U_{2j}(\cos\theta)\,,
\nn
\ee
where $U_{2j}$  is the Chebyshev polynomial of the second kind of degree $2j$.
}

We can compute this series directly. To this purpose, we introduce ``dual spins'' as in \cite{Freidel:2013fia}:
\be
k_{3}=j_{1}+j_{2}-j_{3}
\,,\quad
k_{1}=j_{2}+j_{3}-j_{1}
\,,\quad
k_{2}=j_{3}+j_{1}-j_{2}
\,,\quad
k_{1}+k_{2}+k_{3}=j_{1}+j_{2}+j_{3}
\,,
\ee
\be
2j_{1}=k_{2}+k_{3}
\,,\quad
2j_{2}=k_{1}+k_{3}
\,,\quad
2j_{3}=k_{2}+k_{1}
\,,
\ee
and re-write the generating function as:
\be
\cZ_{\Theta}[Y_{1,2,3}]
=
\sum_{k_{1,2,3}\in\N}
(-1)^{k_{1}+k_{2}+k_{3}}
\f{(k_{1}+k_{2}+k_{3}+1)!}{k_{1}!k_{2}!k_{3}!}
(Y_{2}Y_{3})^{k_{1}}(Y_{1}Y_{3})^{k_{2}}(Y_{1}Y_{2})^{k_{3}}
\,.
\ee
This is easily resummed and gives, as expected,  the loop polynomial of the $\Theta$ graph:
\be
\cZ_{\Theta}[Y_{1,2,3}]
=
\f1{(1+Y_{1}Y_{2}+Y_{2}Y_{3}+Y_{1}Y_{3})^2}
=
\f1{P_{\Theta}[Y_{1},Y_{2},Y_{3}]^2}
\,.
\ee
The roots of the loop polynomial, $P_{\Theta}[Y_{1,2,3}]=0$, gives the zeros of the Ising partition function, $\cI_{\Theta}[Y_{1,2,3}]=0$,  and corresponds to the poles of the spin network generating function $\cZ_{\Theta}[Y_{1,2,3}]\rightarrow\infty$.

\subsection{Dual generating function on triangle graph}

The graph dual to the $\Theta$ graph is the triangle graph $\Delta$, as drawn on fig.\ref{fig:Theta-triangle}. The triangle graph is not 3-valent, but the generating function for spin networks can still be defined and computed with similar formulas. To make sure we work with consistent conventions, it is simpler to start from the loop polynomial on the triangle graph, $P_{\Delta}[Y^*_{1,2,3}]=1+Y_{1}^*Y_{2}^*Y_{3}^*$ and to work out the corresponding weights for the generating function of the spin network evaluation by simply expanding the squared inverse.

Another way to proceed would be to start with a larger 3-valent graph containing the triangle graph as a subgraph and derive the generating function for spin network evaluations on $\Delta$ from the loop polynomial on the larger graph by setting the extra edge couplings to 0. We work this out explicitly, showing how to derive the loop polynomial $P_{\Delta}$ from the loop polynomial $P_{T}$ on the tetrahedron graph $T$ in section \ref{graphreduction}.

Here, on the graph $\Delta$, the three spins $j_{1,2,3}$ on the three edges around the loop must necessarily be equal, $j_{1}=j_{2}=j_{3}=j$ and we expand the spin network generating function as:
\be
\cZ_{\Delta}[Y^*_{1,2,3}]
=
\f1{P_{\Delta}[Y^*_{1,2,3}]^2}
=
\f1{(1+Y_{1}^*Y_{2}^*Y_{3}^*)^2}
=
\sum_{j\in\f\N2} (-1)^{2j}(2j+1)
(Y_{1}^*Y_{2}^*Y_{3}^*)^{2j}
\,.
\ee
Then the duality formula for  the Ising partition functions reads, with $Y^*_{i}=\cD(Y_{i})$:
\be
P_{\Theta}[Y_{1,2,3}]
=
\f{3\,P_{\Delta}[Y^*_{1,2,3}]}{\prod_{i}^3(1+Y^*_{i})}
\,.
\ee
This means that  there is a one-to-one correspondence in the complex plane between the roots of the loop polynomial $P_{\Theta}[Y_{1,2,3}]$ on the $\Theta$ graph and the loop polynomial $P_{\Delta}[Y^*_{1,2,3}]$ on the dual graph.

\subsection{Roots from Triangle Geometry with Complex Edge Lengths}

Let us look for the roots of the loop polynomial on the $\Theta$ graph,  $P_{\Theta}[Y_{1},Y_{2},Y_{3}]=0$, which are also the Fisher zeros for the Ising partition function and the poles of the Ponzano-Regge amplitude for that boundary graph. The obvious method is to compute $Y_{3}$ from $Y_{1}$ and $Y_{2}$:
\be
P_{\Theta}[Y_{1},Y_{2},Y_{3}]=0
\quad\Longleftrightarrow\quad
Y_{3}=-\f{1+Y_{1}Y_{2}}{Y_{1}+Y_{2}}
\,,
\ee
as long as $Y_{1}+Y_{2}\ne 0$. If $Y_{1}+Y_{2}= 0$ vanishes, the variable $Y_{3}$ is irrelevant and can take arbitrary complex values while it is enough to have $1+Y_{1}Y_{2}=0$, i.e. $Y_{1}=\pm 1=-Y_{2}$. Not only this method breaks the symmetry between the three edge variables, but it does not carry any geometrical interpretation.

Here we are interested in providing this equation and its solutions with a geometrical interpretation in terms of triangle geometry. Indeed, the spin network evaluation sums over configurations of three spins $j_{1,2,3}$ defining a triangular configuration with triangular inequalities, $|j_{2}-j_{3}|\le j_{1}\le (j_{2}+j_{3})$ and so on. Furthermore, as was explained in \cite{Bonzom:2015ova}, the generating function for spin network evaluations is a sum over geometric configurations up to a global scale factor. Thus we expect a geometrical interpretation in terms of triangles up to global scale. We therefore expect roots of the loop polynomial $P_{\Theta}$ to be parameterized  in terms of triangle angles.

\subsubsection{Geometric solutions}

\begin{figure}[h!]

\begin{tikzpicture}[x=0.75pt,y=0.75pt,yscale=-0.7,xscale=0.7]

\coordinate(A) at  (321.5,78);
\coordinate(B) at  (200,217) ;
\coordinate(C) at  (411.5,217);

\draw    (A) -- (B) node[midway,left]{$l_{3}$};
\draw    (B) -- (C) node[midway,below]{$l_{1}$};
\draw    (C) -- (A) node[midway,right]{$l_{2}$};

\pic [draw, "$\phi_1$", angle eccentricity=1.5] {angle = B--A--C};
\pic [draw, "$\phi_2$", angle eccentricity=1.5] {angle = C--B--A};
\pic [draw, "$\phi_3$", angle eccentricity=1.5] {angle = A--C--B};

\end{tikzpicture}

\caption{Triangle geometry: angles and edge lengths.
\label{fig:triangle}}

\end{figure}
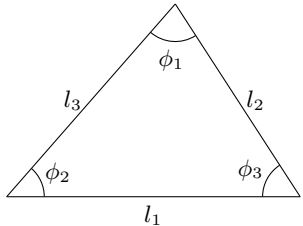

One can easily find solutions of $P_\Theta[Y_{1,2,3}] =0$, or equivalently to $P_\Delta[Y^*_{1,2,3}]=0$, using the fact that the angles of a triangle sum to $\pi$. Consider a triangle as drawn in fig.\ref{fig:triangle}.

Looking for roots of $P_{\Delta}$ with unit norm  (i.e. phases), it is straightforward to see that they are simply given by three angles\footnotemark{} summing to $\pi$. A solution is therefore 
\be
Y_a^* = e^{i\phi_a} \qquad\Rightarrow\quad
P_{\Delta}[e^{i\phi_1},e^{i\phi_2},e^{i\phi_3}]=1+e^{i(\phi_1+\phi_2+\phi_3)}=0
\,.
\ee
\footnotetext{
We could also use the angles of the contact triangle,
\begin{equation}
\hat{\phi_a}=\f{\phi_b+\phi_c}2 \qquad \text{for $\{a,b,c\}=\{1,2,3\}$},
\nn
\end{equation}
defined by the three points of contact of the triangle's incircle with its edges. Similarly, the angles of any special triangle within the original triangle can in fact be used.
}
We  apply the duality map $\mathcal{D}$ to find the dual roots, i.e. the roots of $P_{\Theta}$:
\begin{equation}
Y_a = \mathcal{D}(Y^*_a) = \frac{1 - e^{i\phi_a}}{1 + e^{i\phi_a}} = -i \tan \frac{\phi_a}{2}\,.
\end{equation}
Expanding the trigonometric identity $\cos \frac{1}{2}(\phi_1+\phi_2+\phi_3) = 0$, it can indeed be directly checked that\footnotemark{}:
\begin{equation} \label{TanPhi/2}
P_{\Theta}\bigg[\Big\{-i \tan \frac{\phi_a}{2}\Big\}_{a}\bigg]
=
1 - \tan \frac{\phi_1}{2}\, \tan \frac{\phi_2}{2} - \tan \frac{\phi_2}{2}\,\tan \frac{\phi_3}{2} - \tan \frac{\phi_3}{2}\,\tan \frac{\phi_1}{2} = 0
\,.
\end{equation}
\footnotetext{
Equivalent to \eqref{TanPhi/2}, we can use the law of tangents within a triangle, obtained by expanding $\sin (\phi_1+\phi_2+\phi_3) = 0$,
\be
\f{1}{\tan\phi_1\,\tan\phi_2}+\f{1}{\tan\phi_2\,\tan\phi_3}+\f{1}{\tan\phi_3\,\tan\phi_1}=1
\,.
\nn
\ee
This leads to roots of $P_{\Theta}$, which are again purely imaginary and thus roots of $P_\Delta$ which are phases, as follows
\be
\label{circumangles}
Y_a = \frac{i}{\tan\phi_a} \qquad \Leftrightarrow \qquad Y^*_a = -e^{2i\phi_a}
\quad\Rightarrow\quad
P_{\Delta}[Y^*_{1,2,3}]= Y^*_{1}Y^*_{2}Y^*_{3}+1 =0
\,.
\nn
\ee
Twice the triangle angles actually correspond to the angles around the center of the circumscribed circle. 
}
Notice that purely imaginary roots of $P_\Theta$ are equivalent to roots of $P_\Delta$ of unit norm under the action of $\mathcal{D}$,
\be
Y\in i\R\Longleftrightarrow |Y^*|=|\cD(Y)|=1\,, \qquad Y=\f{-i}{\tan\f\phi2}
\Longleftrightarrow
Y^*=e^{i\phi}
\,.
\ee
The above solutions  provide a two-real-parameter family of zeros of the loop polynomials $P_{\Delta}$ and $P_{\Theta}$. 
We will see below that all the roots of $P_\Theta$ can be parametrized this way by a continuation of the triangle geometry to the complex plane, i.e. using complex edge lengths instead of just real ones.

\subsubsection{Parametrizing all solutions using complex edge lengths} \label{sec:TriangleCycleVariables}

Here we use a method that we will use again in the more complicated case of the tetrahedron graph in Section \ref{sec:FisherTet}. 

First we introduce  cycle variables $L_{ab}$ defined by
\begin{equation}
L_{ab} = Y_a\,Y_b\,,
\end{equation}
for any $\{a,b\}\subset \{1,2,3\}$. It allows to recover $Y_a$ as $Y_a^2 = {L_{ab}\,L_{ac}}/{L_{bc}}$ for $\{a,b,c\}=\{1,2,3\}$ (i.e. $a,b,c$ all different from each other). The equation $P_\Theta[Y_{1,2,3}]=0$ then simply becomes
\begin{equation}
1 + L_{12} + L_{23} + L_{13} = 0,
\end{equation}
which can be solved easily using \emph{homogeneous variables} $\{M_{ab}\}$,
\begin{equation}
L_{ab} = - \frac{M_{ab}}{M_\Theta} \qquad \text{with $M_\Theta = \sum_{\{a,b\}\subset\{1,2,3\}} M_{ab}$}.
\end{equation}
The roots become
\begin{equation}
Y_a^2 = -\frac{M_{ab}\,M_{ac}}{M_{bc}\,M_\Theta}\,.
\end{equation}
One can then introduce the \emph{complex lengths} $l_1, l_2, l_3$ as follows
\begin{equation}
l_a = M_{ab} + M_{ac} \qquad \Leftrightarrow \qquad M_{ab} = \frac{1}{2}(l_a + l_b - l_c).
\end{equation}
It is convenient to introduce the \emph{complex semi-perimeter} $s = \frac12 (l_1 + l_2 + l_3)$ so that
\begin{equation} \label{TriangleCouplings}
Y_a = i \epsilon\,\sqrt{\frac{(s-l_a)\,(s-l_b)}{s\,(s-l_c)}}\,,
\end{equation}
with an overall sign $\epsilon=\pm$. This provides a parametrization of the whole set of complex roots of $P_{\Theta}$ and, by duality, of $P_{\Delta}$.

If we now assume  that the edge lengths are real, $l_a\in\mathbb{R}$, and satisfy the triangle inequalities $|l_b-l_c|\leq l_a \leq l_b + l_c$, then one recognizes \eqref{TriangleCouplings} as the tangent of the half-angles:
\begin{equation}
Y_a = i\epsilon\, \sqrt{\frac{(s-l_a)\,(s-l_b)}{s\,(s-l_c)}} = i \epsilon \tan \frac{\phi_a}{2}\,.
\end{equation}
where $s$ is the (real) semi-perimeter of the triangle. Indeed, there are simple, well-known formula for both the cosine and sine of the angle $\phi$ from which we find $\tan \phi/2$,
\begin{equation} \label{CosSin}
\tan\f{\phi_a}2
=
\f{1-\cos\phi_a}{\sin\phi_a}
\,, \qquad
\cos \phi_a = \frac{l_b^2 + l_c^2 - l_a^2}{2\,l_b\,l_c}
\,, \qquad
\sin\phi_a = \frac{2\mathcal{A}}{l_b\,l_c}
\,,
\end{equation}
where the triangle area $\mathcal{A}$ is the real positive root of the Heron formula
\begin{equation} \label{Heron}
\cA^2 = s(s-l_{1})(s-l_{2})(s-l_{3}) = \f1{4^2}(l_{1}+l_{2}+l_{3})(-l_{1}+l_{2}+l_{3})(l_{1}-l_{2}+l_{3})(l_{1}+l_{2}-l_{3}).
\end{equation}
Notice that the freedom in $\epsilon=\pm 1$ corresponds to the possibility of taking the positive or negative root of the Heron formula, thus flipping the sign of $\sin\phi_a$ and hence of $\tan \phi_a/2$.

All the above formula can be directly continued in the complex case. In particular, the Heron formula has two complex roots $\mathcal{A}_\pm = \epsilon \sqrt{s(s-l_{1})(s-l_{2})(s-l_{3})}$ and one can extends \eqref{CosSin} to complex lengths. It comes that the value of $Y_a$ given in \eqref{TriangleCouplings} really is a continuation of $i\tan\phi_a/2$.

Applying $\mathcal{D}$, one finds
\begin{equation}
Y_{a}^* = \frac{1-i\epsilon\tan(\phi_a/2)}{1+i\epsilon\tan(\phi_a/2)} = \exp i\epsilon\phi_a = \f{l_{b}^2+l_{c}^2-l_{a}^2+4i\cA_{\pm}}{2l_{b}l_{c}}\,.
\end{equation}

Notice the roots as a function of the three complex edge lengths is invariant under arbitrary complex rescaling of the complex edge lengths, $l_{e}\rightarrow \lambda l_{e}$ with $\lambda\in\C$. This implies that it in fact depends on two complex parameters, i.e. four real parameters as wanted.

\subsection{Cevian Parametrization of the Loop Polynomial's Roots}
\label{cevian}

An alternative way to get all complex roots of $P_{\Delta}$ and $P_{\Theta}$ without resorting to complexifying the triangle geometry is to keep a real triangle but add the extra data of a point in the plane, as explained below.

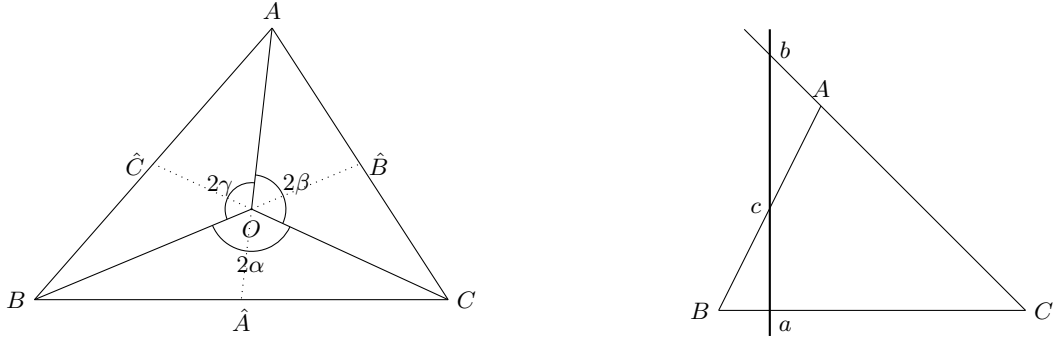
\begin{figure}[h!]

\begin{subfigure}[t]{.4\linewidth}
\centering
\begin{tikzpicture}[scale=0.026]

\coordinate(A) at  (321.5,-78);
\coordinate(a) at  (305.75,-217);
\coordinate(B) at  (200,-217) ;
\coordinate(b) at  (366.5,-147.5) ;
\coordinate(C) at  (411.5,-217);
\coordinate(c) at  (260.75,-147.5);

\coordinate(O) at  (311,-170.66);
\draw (O)++(0,-10) node{$O$};

\draw (A) node[above]{$A$};
\draw (a) node[below]{$\hat{A}$};
\draw (B) node[left]{$B$};
\draw (b) node[right]{$\hat{B}$};
\draw (C) node[right]{$C$};
\draw (c) node[left]{$\hat{C}$};

\draw    (A) -- (B) ;
\draw    (B) -- (C) ;
\draw    (C) -- (A) ;
\draw    (A) -- (O) ;
\draw    (B) -- (O) ;
\draw    (C) -- (O) ;
\draw[dotted]    (O) -- (a) ;
\draw[dotted]    (O) -- (b) ;
\draw[dotted]    (O) -- (c) ;

\pic [draw, "$2\alpha$", angle eccentricity=1.3,angle radius=16] {angle = B--O--C};
\pic [draw, "$2\beta$", angle eccentricity=1.5,angle radius=13] {angle = C--O--A};
\pic [draw, "$2\gamma$", angle eccentricity=1.5,angle radius=10] {angle = A--O--B};

\end{tikzpicture}

\caption{
Considering the triangle $(ABC)$ and a point $O$, the Cevians are the lines linking the three vertices of the triangle to the point $O$. We call $\hat{A}$, $\hat{B}$, $\hat{C}$, their intersection points with the triangle edges. The length ratios satisfy the relation
$
\f{{B\hat{A}}}{{\hat{A}C}}\,
\f{{C\hat{B}}}{{\hat{B}A}}\,
\f{{A\hat{C}}}{{\hat{C}B}}
=1
$.
\label{fig:cevian}
}

\end{subfigure}
\hspace*{10mm}
\begin{subfigure}[t]{.4\linewidth}
\begin{tikzpicture}[scale=0.68]

\coordinate(A) at  (0,4);
\coordinate(a) at  (-1,0);
\coordinate(B) at  (-2,0) ;
\coordinate(b) at  (-1,5) ;
\coordinate(C) at  (4,0);
\coordinate(c) at  (-1,2);

\draw (A) node[above]{$A$};
\draw (a)++(0.3,-0.3) node{$a$};
\draw (B) node[left]{$B$};
\draw (b)++(0.3,0.1) node{$b$};
\draw (C) node[right]{$C$};
\draw (c) node[left]{$c$};

\draw    (A) -- (B) ;
\draw    (B) -- (C) ;
\draw    (C) -- (A) ;
\draw[thick]   (-1,5.5) -- (-1,-0.5);
\draw (A)-- (-1.5,5.5);

\end{tikzpicture}

\caption{
Considering the triangle $(ABC)$ and an arbitrary line in the plane, we call $a$, $b$, $c$, their intersection points of the line with the triangle edges. Then the oriented length ratios satisfy the Menela\"us theoreom
$
\f{{Ba}}{{aC}}\,
\f{{Cb}}{{bA}}\,
\f{{Ac}}{{cB}}
=-1
$.
\label{fig:menelaus}}

\end{subfigure}

\caption{
Ceva and Menela\"us theoreom for ratios in a triangle.
}

\end{figure}

Coming back to the roots of the polynomial $P_{\Delta}$ in terms of the double triangle angles or equivalently in terms of the angles around the center of the circumscribed circle given in \eqref{circumangles}, we could choose any point $O$ on the plane and use the angles  between the lines linking $O$ to the three points of the triangle, $2\gamma=\widehat{AOB}$, $2\alpha=\widehat{BOC}$, $2\beta=\widehat{COA}$, as drawn on fig.\ref{fig:cevian}. This allows to define more general complex solutions based on the Cevians going through the point $O$. Indeed, Ceva theorem\footnotemark{} implies that the product of the ratio of the oriented length of the split segments is equal to 1:
\be
\f{{B\hat{A}}}{{\hat{A}C}}\
\f{{C\hat{B}}}{{\hat{B}A}}\
\f{{A\hat{C}}}{{\hat{C}B}}
=1
\,.
\ee
\footnotetext{
Ceva theorem is easily understood in terms of triangle areas, which interprets the length ratios as ratios of the areas of the three triangles with $O$ as summit. Assuming that $O$ is inside the triangle so that the signs are all positive for the sake of simplicity, we have:
\be
\f{{B\hat{A}}}{{\hat{A}C}}
=
\f{\cA_{OB\hat{A}}}{\cA_{O\hat{A}C}}
=
\f{\cA_{AB\hat{A}}}{\cA_{A\hat{A}C}}
=
\f{\cA_{OAB}}{\cA_{OAC}}
\,,\qquad
\f{{B\hat{A}}}{{\hat{A}C}}\,
\f{{C\hat{B}}}{{\hat{B}A}}\,
\f{{A\hat{C}}}{{\hat{C}B}}
=
\f{\cA_{OAB}}{\cA_{OAC}}
\f{\cA_{OCB}}{\cA_{OBA}}
\f{\cA_{OAC}}{\cA_{OCB}}
=1
\,.
\nn
\ee
}
Upon switching signs, this gives real roots of $P_{\Delta}$, which once combined with the phases provides a complete parametrization in terms of Cevians of the complex roots of $P_{\Delta}$:
\be
\label{Ceva}
Y_{1}^*=-\f{{B\hat{A}}}{{\hat{A}C}}
\,,\,\,
Y_{2}^*=-\f{{C\hat{B}}}{{\hat{B}A}}
\,,\,\,
Y_{3}^*=-\f{{A\hat{C}}}{{\hat{C}B}}
\qquad
\Rightarrow\quad
P_{\Delta}[Y^*_{1,2,3}]=0
\,,
\ee
\be
Y_{1}^*=-e^{2i\alpha}\,\f{{B\hat{A}}}{{\hat{A}C}}
\,,\,\,
Y_{2}^*=-e^{2i\beta}\,\f{{C\hat{B}}}{{\hat{B}A}}
\,,\,\,
Y_{3}^*=-e^{2i\gamma}\,\f{{A\hat{C}}}{{\hat{C}B}}
\qquad
\Rightarrow\quad
P_{\Delta}[Y^*_{1,2,3}]=0
\,.
\ee
Indeed, given values $Y^*_{1,2,3}$ satisfying this ansatz $P_{\Delta}[Y^*_{1,2,3}]=0$, we choose $O$ as the origin of the plane, define the three Cevians going through $O$ with the correct angles $\alpha,\beta,\gamma$ then identify the points $A,B,C$ along those lines in order to realize the values $Y^*_{1,2,3}$ through the ansatz above. Moreover, a simple counting checks that this ansatz depends on four real parameters as needed.

A variation on this ansatz uses a line instead of the point $O$ and is based on Menela\"us theorem, as illustrated  fig.\ref{fig:menelaus}. Actually, since roots $P_{\Delta}$ simply involve three angles summing to $\pi$ and three real numbers whose product gives 1,
\be
P_{\Delta}[Y^*_{1,2,3}]=0
\quad\Longleftrightarrow\quad
Y^*_{i}=\rho_{i}e^{i\theta_{i}}
\quad\textrm{with}\,\,
\left|
\begin{array}{l}
\rho_{1}\rho_{2}\rho_{3}=1 \\
\theta_{1}+\theta_{2}+\theta_{3}=\pi
\end{array}
\right.
\,,
\ee
where the $\rho$'s are the modulus of the roots and the $\theta$'s their arguments, there are actually plenty of possible ways to generate such solutions from a triangle geometry. The important point is that we can not use one triangle to generate both phases and modulus. Indeed, a triangle $(ABC)$ up to scale is determined by two angles and this would be enough to generate a two-parameter family of complex roots, such as only the phase $e^{i\theta_{i}}$ or simply the signed modulus $-\rho_{i}$. So adding an extra point $O$ apart from the triangle $(ABC)$ allows to introduce two extra parameters in order to provide a complete parametrization of the complex roots of $P_{\Delta}$.

For instance, if we choose as angles $\theta$'s the angles $(\alpha,\beta,\gamma)$ around $O$ then the modulus need to depend on a choice of triangle $(ABC)$, such as the ratio of triangle areas as in the Cevian ansatz  above or the ratio of the triangle edge lengths $(\f bc,\f ca,\f ab)$ (which are actually the ratio of the sine of the triangle angles and the ratio of the triangle areas taking as internal summit the center of the incircle of the triangle). On the other hand, if we choose as angles $\theta$'s the triangle angles $(\phi_1,\phi_2,\phi_3)$ then the modulus need to depend explicitly on the choice of the point $O$, such as the ratio $(\f{OB}{OC},\f{OC}{OA},\f{OA}{OB})$ of the distances between $O$ and the points $(ABC)$.

Then taking the dual of these solutions by $\cD$ provides a complete parametrization of the roots of the loop polynomial  $P_{\Theta}$.
In particular, considering  the real roots \eqref{Ceva} of $P_{\Delta}$ gives purely real roots of  $P_{\Theta}$.
It would be interesting to investigate if there is a deeper relation between the Cevian ansatz for real triangles and the complex triangle ansatz using complex edge lengths  derived in the previous section.

We now turn to the main object of interest of the present work, that is the tetrahedron graph.

\section{6j Generating function and Self-duality}
\label{sec:tetra}

\begin{figure}[h!]
\begin{subfigure}[t]{.4\textwidth}
\includegraphics[scale=.5]{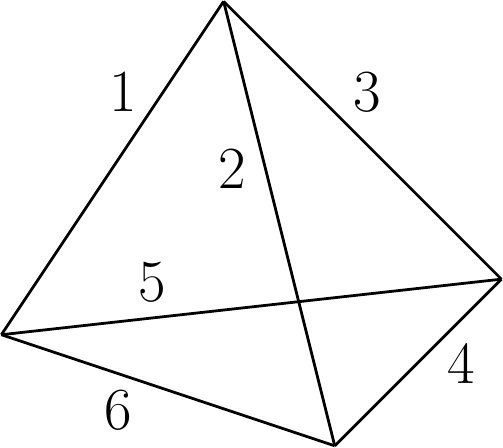}
\caption{\label{fig:6jGraph} The tetrahedron graph $T$ with labeled edges used to define the 6j-symbol.}
\end{subfigure}
\hspace{1cm}
\begin{subfigure}[t]{.4\textwidth}
\includegraphics[scale=.5]{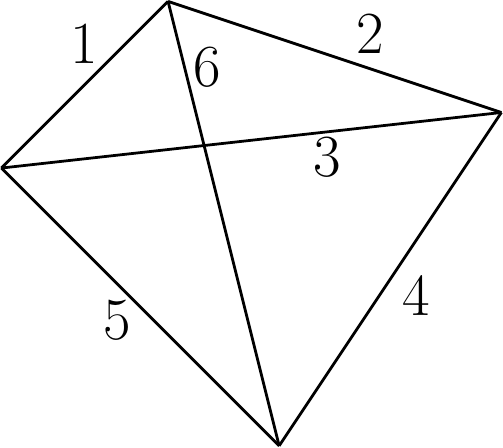}
\caption{\label{fig:6jDualGraph} The dual tetrahedron $T^*$ can be obtained mapping vertices of $T$ to triangles of $T^*$ and triangles of $T$ to vertices of $T^*$. This is equivalent to switching opposite edges of $T$, $1\leftrightarrow 4$, $2\leftrightarrow 5$, $3\leftrightarrow 6$.
}
\end{subfigure}
\caption{\label{fig:6j} The tetrahedron graph $T$ and its dual tetrahedron graph $T^*$.}
\end{figure}

Let us now consider the tetrahedron graph $T$, with four vertices and six edges labeled $e_1,.., e_6$ as shown in fig.\ref{fig:6jGraph}. The spin network evaluation on $T$ depends on six spins $j_{1,..,6}$ living on the edges and consists in the contraction of the four 3j-symbols corresponding to the graph's four vertices. This gives Wigner's 6j-symbol $\left\{\begin{smallmatrix} j_1 & j_2 & j_3\\ j_4 & j_5 & j_6 \end{smallmatrix}\right\}$.

In this section, we will study the spin network generating function on the tetrahedron graph $T$, i.e. the generating function for 6j-symbols, and we will apply the low/high temperature duality of the Ising model to  relate it to the spin network generating function on the dual graph $T^*$.
The dual graph $T^*$ is actually also a tetrahedral graph, but it inherits a different edge labeling, as illustrated on fig.\ref{fig:6jDualGraph}, since vertices of $T$ become triangles in $T^*$ and the other way around.
The low/high temperature duality of the Ising model then leads to a new self-duality formula for the generating function of the 6j-symbols and thus for the 6j-symbols themselves.

\subsection{The 6j Generating function and tetrahedron loop polynomial}
\label{sec:6jdef}

The generating functional for the 6j-symbols on the tetrahedron graph $T$ is defined as:
\be
\cZ_{T}[\{Y_e\}]
\,=\,
\sum_{\{j_{e\}}} \prod_{v}\Delta_v(j_e)\,
\sixj{j_{1}}{j_{2}}{j_{3}}{j_{4}}{j_{5}}{j_{6}} \,\prod_{e}Y_{e}^{2j_{e}}
\,,
\ee
where the triangle coefficients $\Delta_v(a,b,c)$ for three spins $a, b, c$ meeting at a vertex $v$ of $T$ are defined as:
\begin{equation}
\Delta_v(a,b,c) = \sqrt{\frac{(a+b+c+1)!}{(a+b-c)!(a-b+c)!(-a+b+c)!}}
\,.
\end{equation}

It was shown to be exactly computable as a rational function, first by Schwinger \cite{Schwinger} and then by Bargmann \cite{Bargmann},
\be
\cZ_{T}[\{Y_e\}]
\,=\,
\f1{P_{T}[\{Y_e\}]^{2}}
\,,
\qquad
P_{T}[\{Y_e\}]=\sum_{C\subset T} \prod_{e\in C} Y_{e}
\,,
\ee
where $C$ runs through all cycles (or loops) of the tetrahedral graph $T$.
This is the prototype formula for the more general duality formula \eqref{westbury} by Westbury expressing the generating function for spin network evaluations on an arbitrary graph in terms of the corresponding loop polynomial and thereby relating it to the 2D Ising model.

The loop polynomial $P_{T}$ on the tetrahedron graph can be written explicitly:
\be
P_{T}[\{Y_e\}]
=
1
+Y_{1}Y_{2}Y_{6}+Y_{1}Y_{3}Y_{5}+Y_{2}Y_{3}Y_{4}+Y_{4}Y_{5}Y_{6}
+Y_{1}Y_{4}Y_{2}Y_{5}+Y_{2}Y_{5}Y_{3}Y_{6}+Y_{1}Y_{4}Y_{3}Y_{6}
\ee
There are two types of cycles $C$: either it is a \emph{3-cycle}, i.e. a triangle of $T$, and there are four of them, or its is a \emph{4-cycle} which runs through all edges but a pair of opposite ones, and there are three of them. Notice that in the dual graph, the 3-cycles are mapped to vertices of $T^*$, while each 4-cycle remains the same.

\subsection{Scissor Symmetry and Invariance of the 6j Generating Function}

The 6j-symbol is invariant under the Regge symmetries of the tetrahedron. These include  permutations of the 6 spins $j_{1,2,3,4,5,6}$ compatible with the combinatorics of the tetrahedron graph, which thus leave invariant the generating function $\cZ_{T}$ and the corresponding loop polynomial $P_{T}$, but they also contain another type of symmetries, called scissor symmetries, which are generated by the following transformation:
\be
\sixj{j_{1}}{j_{2}}{j_{3}}{j_{4}}{j_{5}}{j_{6}} 
=
\sixj{j_{1}}{\f12(-j_{2}+j_{3}+j_{5}+j_{6})}{\f12(j_{2}-j_{3}+j_{5}+j_{6})}{j_{4}}{\f12(j_{2}+j_{3}-j_{5}+j_{6})}{\f12(j_{2}+j_{3}+j_{5}-j_{6})}
\,.
\ee
This translates into an intriguing invariance of the loop polynomial, if we perform an inversion of the couplings $Y_{2,3,5,6}$ around the 4-cycle $(2356)$ by the corresponding loop monomial $Y_{2}Y_{3}Y_{5}Y_{6}$:
\be
P_{T}[\{Y_{e}\}]=P_{T}[\{\tY_{e}\}]
\ee
with:
\be
\tY_{1}=Y_{1}\,,\quad\tY_{4}=Y_{4}
\,,\quad
\tY_{2}=\f{\sqrt{Y_{2}Y_{3}Y_{5}Y_{6}}}{Y_{2}}
\,,\quad
\tY_{3}=\f{\sqrt{Y_{2}Y_{3}Y_{5}Y_{6}}}{Y_{3}}
\,,\quad
\tY_{5}=\f{\sqrt{Y_{2}Y_{3}Y_{5}Y_{6}}}{Y_{5}}
\,,\quad
\tY_{6}=\f{\sqrt{Y_{2}Y_{3}Y_{5}Y_{6}}}{Y_{6}}
\,.
\ee
This transformation leads to a permutation of monomials corresponding to the loops on the tetrahedron graph, due to the identities $\tY_{2}\tY_{3}=Y_{5}Y_{6}$ and so on for the 6 pairs of edges around the 4-cycle.

\subsection{3-1 Pachner move and Reduction to the Triangle}
\label{graphreduction}

One can contract the tetrahedron graph to get either the triangle graph $\Delta$ or the $\Theta$ graph.
The simpler case is to recover the $\Delta$ graph by erasing the links $1,2,3$ from the tetrahedron graph. This is done by setting the corresponding edge couplings to 0:
\be
P_{T}[Y_{1,2,3}=0,Y_{4},Y_{5},Y_{6}]
=
1+Y_{4}Y_{5}Y_{6}
=
P_{\Delta}[Y_{4},Y_{5},Y_{6}]
\,.
\ee

\begin{figure}[h!]

\centering

\begin{tikzpicture}[scale=0.75]

\coordinate(A) at (0,0);
\coordinate(B) at (-1,-1.7);
\coordinate(C) at (1,-1.7);
\coordinate(a) at (0,1);
\coordinate(b) at (-1.8,-2.5);
\coordinate(c) at (1.8,-2.5);

\draw(A) node{$\bullet$};
\draw(B) node{$\bullet$};
\draw(C) node{$\bullet$};

\draw (A)--(B) node[midway, left]{$Y_{6}$};
\draw (B)--(C) node[midway, below]{$Y_{4}$};
\draw (C)--(A) node[midway, right]{$Y_{5}$};

\draw(A)--(a) node[midway, left]{$Y_{1}$};
\draw(B)--(b)node[near end, right]{$Y_{2}$};
\draw(C)--(c)node[near end, left]{$Y_{3}$};

\draw[thick,<->,>=latex] (2.5,-0.8)--(3.5,-0.8);

\coordinate(O) at (5.5,-0.8);
\draw(O) node{$\bullet$};

\coordinate(x) at (5.5,1);
\coordinate(y) at (3.7,-2.5);
\coordinate(z) at (7.3,-2.5);

\draw(O)--(x) node[midway, left]{$\tY_{1}$};
\draw(O)--(y)node[midway,right]{$\tY_{2}$};
\draw(O)--(z)node[midway, left]{$\tY_{3}$};

\end{tikzpicture}

\caption{The 3$\leftrightarrow$1 Pachner move on a 3-valent graph amounts to blowing up one vertex into a little triangle, or vice-versa. It is possible to relate the loop polynomials of two graphs related by such a move.} 
\label{fig:Pachner31}

\end{figure}
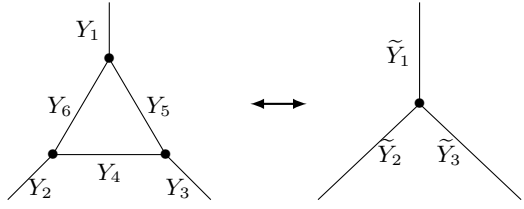

In order to reduce the tetrahedron graph $T$ to the $\Theta$ graph, the procedure is slightly more complicated and we have to contract the cycle $(4,5,6)$ to a single point. This is a 3-1 Pachner move, which can be defined on arbitrary 3-valent graphs. Let us consider an initial graph $\Gamma$, choose a vertex $v_{0}$ on the graph and blow it up into a little loop linking the three edges attached to that vertex, as shown on fig.\ref{fig:Pachner31}. We call the final graph $\tGamma$. Then we can relate the two loop polynomials on $\Gamma$ and $\tGamma$.

The loop polynomial $P_{\tGamma}$ is defined as a sum over loop systems or equivalently all unions of disjoint cycles on the graph $\tGamma$. Starting from a loop system on $\Gamma$, we distinguish loops that avoid the vertex $v_{0}$. We raise them to two loop systems on $\tGamma$: the original one and the original one together with the new loop $(456)$, with the new loop contributing a factor $Y_{4}Y_{5}Y_{6}$ to the polynomial. On the other hand, if a loop travels through the vertex $v_{0}$, it follows one of the three paths, $(12)$, $(23)$ or $(31)$. Once raised to the graph $\tGamma$, each of these paths can follow two ways around the loop $(456)$. For instance, the original path $(12)$ can become $(162)$ or $(1542)$ meaning that the factor $Y_{1}Y_{2}$ acquires a factor and becomes $Y_{1}Y_{2}(Y_{6}+Y_{4}Y_{5})$. If we factor the whole polynomial by an overall $(1+Y_{4}Y_{5}Y_{6})$, we can re-absorb the rescaling for the three monomials $Y_{1}Y_{2}$, $Y_{2}Y_{3}$ and $Y_{3}Y_{1}$ by a rescaling of the individuals couplings $Y_{1,2,3}$:
\be
P_{\tGamma}[Y_{1,2,3,4,5,6},\{Y_{e}\}]
=
(1+Y_{4}Y_{5}Y_{6})\,P_{\Gamma}[\tY_{1,2,3},\{Y_{e}\}]
\,,
\ee
where the generic notation $Y_{e}$ stands for the couplings on all the edges $e$ which are not involved in the 3-1 Pachner move, and
with the renormalized couplings defined as:
\be
\tY_{1}=Y_{1}\sqrt{\f{(Y_{5}+Y_{4}Y_{6})(Y_{6}+Y_{4}Y_{5})}{(Y_{4}+Y_{5}Y_{6})(1+Y_{4}Y_{5}Y_{6})}}
\,,\quad
\tY_{2}=Y_{2}\sqrt{\f{(Y_{4}+Y_{5}Y_{6})(Y_{6}+Y_{4}Y_{5})}{(Y_{5}+Y_{4}Y_{6})(1+Y_{4}Y_{5}Y_{6})}}
\,,\quad
\tY_{3}=Y_{3}\sqrt{\f{(Y_{4}+Y_{5}Y_{6})(Y_{5}+Y_{4}Y_{6})}{(Y_{6}+Y_{4}Y_{5})(1+Y_{4}Y_{5}Y_{6})}}
\,.
\nn
\ee
This should be related to the star-triangle relation for the Ising model \cite{Baxter:1982zz}.
The reverse mapping is simpler. If we set the coupling around the new loop $(456)$ to 1, the renormalization of the couplings disappear and we are left with:
\be
P_{\Gamma}[Y_{1,2,3},\{Y_{e}\}]=
\f12\,
P_{\tGamma}[Y_{1,2,3},Y_{4}=Y_{5}=Y_{6}=1,\{Y_{e}\}]
\,.
\ee
We apply these formula to  $\Gamma=\Theta$ and $\tGamma=T$ and get the relations:
\beq
P_{T}[Y_{1,2,3,4,5,6}]
&=&
(1+Y_{4}Y_{5}Y_{6})P_{\Theta}[\tY_{1,2,3}]
\,,
\eeq
\be
2P_{\Theta}[Y_{1,2,3}]=2(1+Y_{1}Y_{2}+Y_{2}Y_{3}+Y_{1}Y_{3})
=
P_{T}[Y_{1,2,3},Y_{4}=Y_{5}=Y_{6}=1]\,.
\ee
Let us point out that the reduction from the tetrahedron graph $T$ to the triangle graph $\Delta$ involves setting three couplings to 0, while the reduction to the $\Theta$ graph involves setting three couplings to 1. The fact that $\Delta$ is the dual graph of $\Theta$ is consistent with the fact that $0$ is the dual coupling of 1 by the duality map $\cD$.

\subsection{Self-Duality of the 6j Generating Function}

Let us now turn to the duality map on the tetrahedron graph. The low/high temperature duality of the Ising model translates as explained previously into a duality identity \eqref{dualityonP} relating the loop polynomials on $T$ and $T^*$:
\be
\prod_{e}(1+Y^*_{e})\, P_{T}[\{Y_{e}\}]
=
8P_{T^*}[\{Y^*_{e}\}]
\,,\qquad
\prod_{e}(1+Y_{e})P_{T^*}[\{Y^*_{e}\}]
=8P_{T}[\{Y_{e}\}]
\,.
\ee
We can write both the loop polynomial and its dual explicitly:
\be
P_{T}[\{Y_{e}\}]
=
1
+Y_{1}Y_{2}Y_{6}+Y_{1}Y_{3}Y_{5}+Y_{2}Y_{3}Y_{4}+Y_{4}Y_{5}Y_{6}
+Y_{1}Y_{4}Y_{2}Y_{5}+Y_{2}Y_{5}Y_{3}Y_{6}+Y_{1}Y_{4}Y_{3}Y_{6}
\ee
\be
P_{T^*}[\{Y^*_{e}\}]
=
1
+Y^*_{4}Y^*_{5}Y^*_{3}+Y^*_{4}Y^*_{6}Y^*_{2}+Y^*_{5}Y^*_{6}Y^*_{1}+Y^*_{1}Y^*_{2}Y^*_{3}
+Y^*_{1}Y^*_{4}Y^*_{2}Y^*_{5}+Y^*_{2}Y^*_{5}Y^*_{3}Y^*_{6}+Y^*_{1}Y^*_{4}Y^*_{3}Y^*_{6}
\ee
We can check directly the dual relations between these two polynomials when $Y^*_{e}=\cD(Y_{e})$.
We translate these relations to a identity for the generating function of the 6j-symbols, since it is given by the squared inverse of the loop polynomial:
\be
\label{duality1}
\cZ_{T}[\{\tanh y_{e}\}]
\,
\prod_{e}\f1{\cosh^2 y_{e}}
=
8\cZ_{T^*}[\{e^{-2y_{e}}\}]
\,
\prod_{e}e^{-2y_{e}}
\,.
\ee
which reads explicitly once expanded as series over spins as
\begin{equation} \label{Duality6jGF}
\sum_{\{j_e\}} \begin{Bmatrix} j_1& j_2& j_3\\ j_4& j_5& j_6\end{Bmatrix} \prod_v \Delta_v(j_e) \prod_e \frac{(\tanh y_e)^{2j_e}}{(\cosh y_e)^2} = 4^3 \sum_{\{k_e\}} \begin{Bmatrix} k_4& k_5& k_6\\ k_1& k_2& k_3\end{Bmatrix} \prod_{v^*} \Delta_{v^*}(k_e) \prod_e e^{-2(2k_e+1)y_e}
\end{equation}
Notice that it is not the same 6j-symbols on the left and right hand sides, since although the 6j-symbol is invariant under permutations of its columns, it is not invariant under the switch of its two lines. The duality here exchanges opposite edges of the tetrahedron, thereby switching the 3-cycles into the dual 3-cycles.

When $Y_{e}=\tanh y_{e}$ goes 0 (resp. 1) then $Y^*_{e}=e^{-2y_{e}}$ goes to 1 (resp. 0). This means that the role of the large (resp. small) spins on the left hand side is played by the small (resp. large) spins on the right hand side: the low/high temperature duality for the Ising model becomes a UV-IR duality for the spin network geometry.

\subsection{Duality Transform on the 6j-symbol}

\begin{figure}[h!]

\begin{subfigure}[t]{.45\linewidth}
\includegraphics[width=60mm]{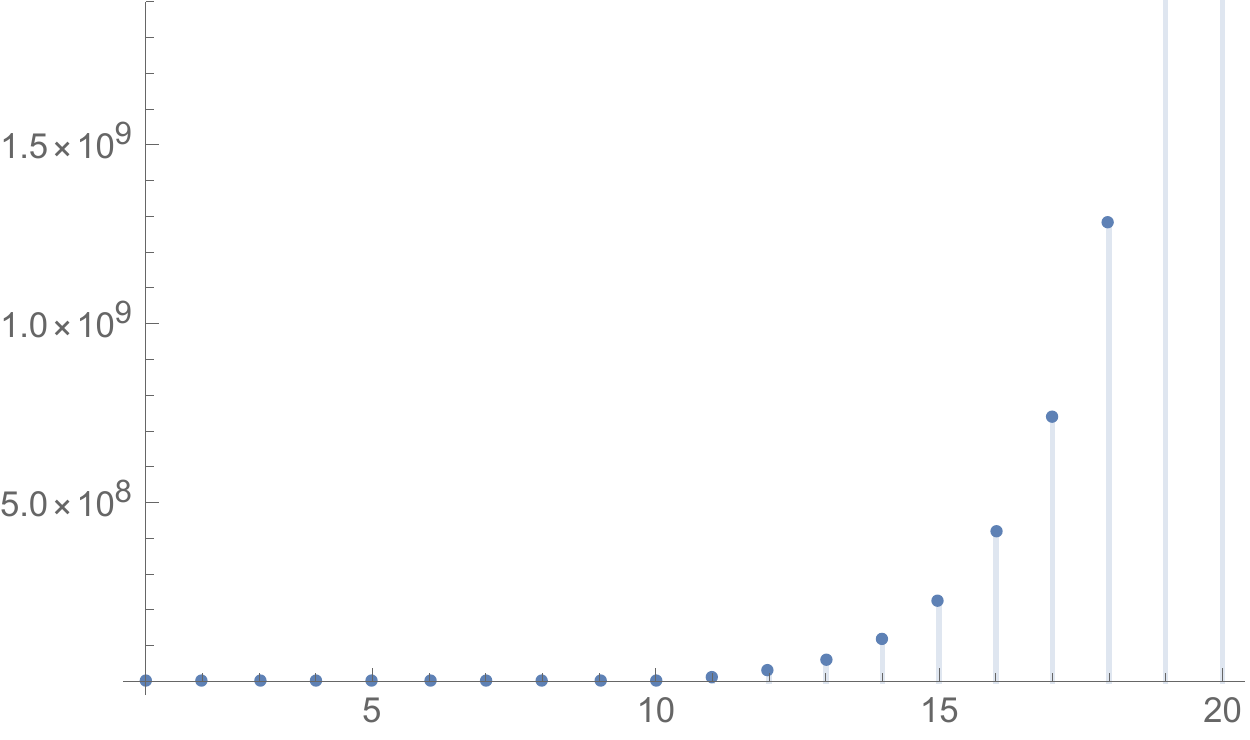}
\caption{
Plot of the figurate number $T(p,q)$ for $q=11$ and $p$ ranging from 1 to 20: the fast growth turns out to be asymptotically a power law.
}
\end{subfigure}
\hspace*{10mm}
\begin{subfigure}[t]{.45\linewidth}
\includegraphics[width=60mm]{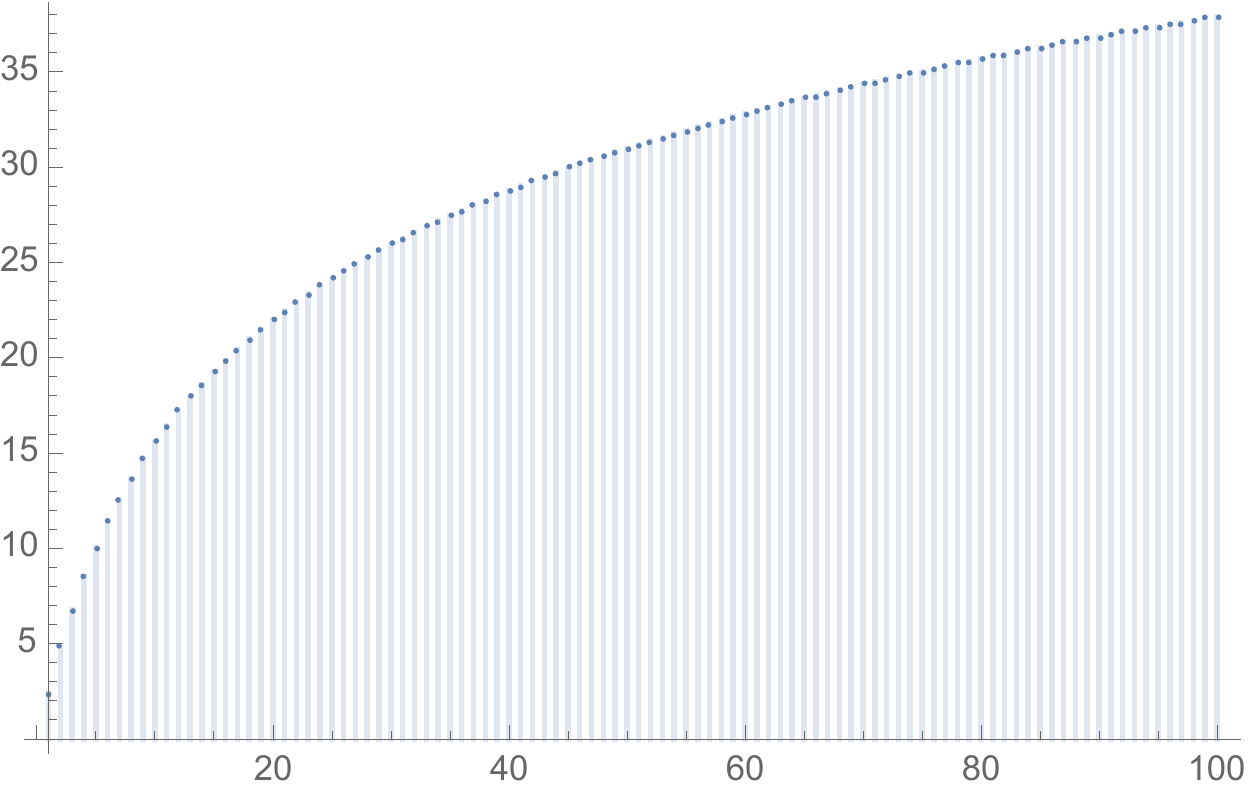}

\caption{
Plot of the log of the figurate number $\ln T(p,q)$ for $q=11$ and $p$ ranging from 1 to 100: the logarithm growth (versus a linear growth) is a signature of a power law.
}

\end{subfigure}

\begin{subfigure}[t]{.45\linewidth}
\includegraphics[width=60mm]{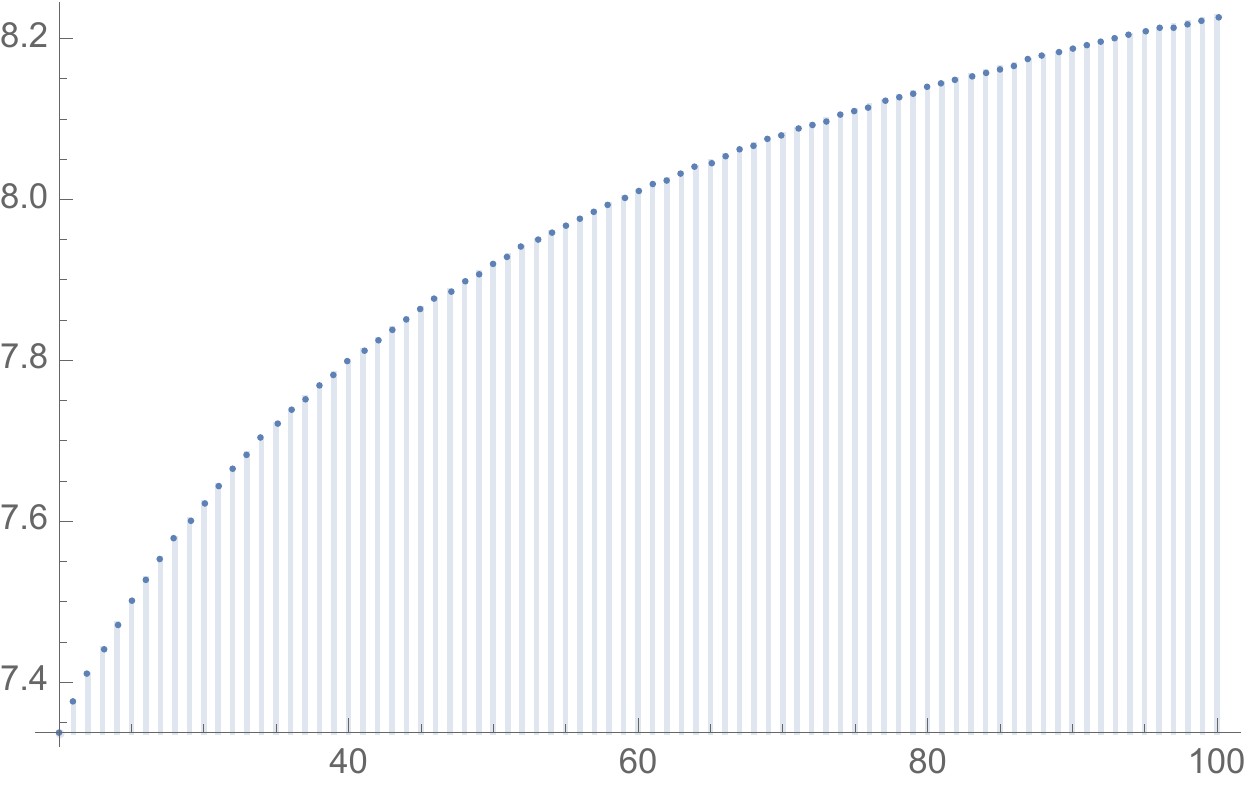}
\caption{
Plot of the log of the ratio  $\ln T(p,q) /\ln p$ for $q=11$ and $p$ ranging from 20 to 100: numerical calculations for large $p$ indicate that the ratio grows slowly towards the limit value $q$ and always stays below.}

\end{subfigure}
\hspace*{10mm}
\begin{subfigure}[t]{.45\linewidth}
\includegraphics[width=60mm]{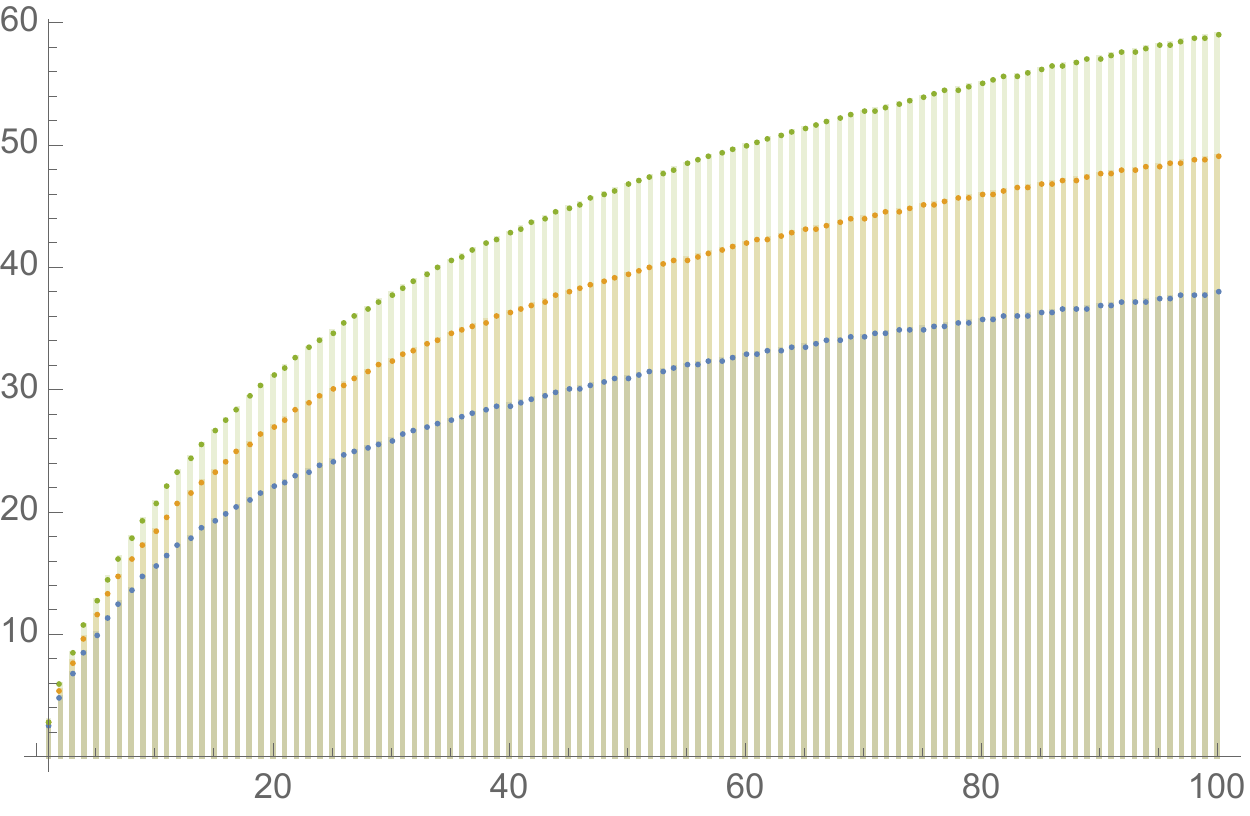}

\caption{
Plot of the log of the figurate number $\ln T(p,q)$ for $p$ ranging from 1 to 100 and various values $q=(11,15,19)$ (from bottom to top): the apparently linear scaling with $q$ is consistent with the crude estimate in $q\ln p$ in the asymptotic limit.}

\end{subfigure}

\caption{
Plots of the figurate number $T(p,q)$ and its logarithm, with $p,q$ integers related to the spins by $(p,q)=(2j+1,2k+1)$: we keep $q$ fixed and look at the evolution of $T(p,q)$ with $p$ running over integer values.
\label{fig:plot-figurate}}

\end{figure}

We can write both sides of eqn \eqref{Duality6jGF} as series in $Y^*_e$ and translate the equalities between the coefficients of both sides as a identity of the 6j-symbols. The right hand side of \eqref{Duality6jGF} is naturally a power series in $Y^*_e = e^{-2y_e}$. As for the left hand side, we write
\begin{equation}
\frac{(\tanh y)^{2j}}{(\cosh y)^2} = 4 Y^* \frac{(1-Y^*)^{2j}}{(1+Y^*)^{2(j+1)}},
\end{equation}
which is known to be the generating function of figurate numbers $(T(2j+1, 2k+1))_{k\in\mathbbm{N}/2}$ for the $(2j+1)$-dimensional cross polytopes (reference \href{http://www.oeis.org/A142978}{A142978 on oeis.org}),
\begin{equation}
Y^* \frac{(1-Y^*)^{2j}}{(1+Y^*)^{2(j+1)}} = \sum_{k\in\mathbbm{N}/2} (-1)^{2k} T(2j+1, 2k+1)\, Y^*{}^{2k+1}.
\end{equation}
Explicitly, the figurate numbers are expressed as sum over product of binomial coefficients,
\be
\label{figuratebinomial}
T(p,q)=\sum_{n=0}^{p-1}\bin{n}{p-1}\bin{p}{q+n}
\,,
\ee
which we plot in fig.\ref{fig:plot-figurate}. These figurate numbers satisfy some interesting properties such as a symmetry under exchange of the arguments, $pT(p,q)=qT(q,p)$ and can be repackaged into a 2-variable generating function:
\be
\sum_{p,q\ge1}T(p,q)x^py^q=
\f{xy}{(1-y)(1-x-y-xy)}
\,.
\ee
Then equating the series in $Y^*$ of the two sides of  \eqref{Duality6jGF} gives:
\begin{equation}
\label{duality2}
2^6 \sum_{\{j_e\}} \begin{Bmatrix} j_1& j_2& j_3\\ j_4& j_5& j_6\end{Bmatrix} \prod_v \Delta_v(j_e) \prod_e (-1)^{2k_{e}} T(2j_e+1, 2k_e+1) = \begin{Bmatrix} k_4& k_5& k_6\\ k_1& k_2& k_3\end{Bmatrix} \prod_{v^*} \Delta_{v^*}(k_e).
\end{equation}
This is our expression for the self-duality of the 6j-symbol. For fixed spins $k$'s, the sum over the spins $j$'s is  unbounded, with an oscillating series due to the 6j-symbol. The more rigorous expression for this duality formula is probably through the re-summation over the spins $k$'s as the identity \eqref{duality1} or \eqref{Duality6jGF} directly on the generating function of the 6j-symbol.

The map from the spins $j_{e}$ to the spins $k_{e}$ is done independently on each edge $e$ using the figurate  number  $T(2j_e+1, 2k_e+1)$ (and the sign $(-1)^{2k_{e}}$). Since a crude estimate\footnotemark{} of the figurate number $T(2j+1,2k+1)$ indicates that it grow as $(2j+1)^{2k+1}$, this {\it figurate transform} can be considered as a discrete equivalent of the Mellin transform and it would be interesting to further investigate its mathematical properties and its potential application to signal analysis.
\footnotetext{The simplest method to get an estimate of the figurate number $T(p,q)$ for a priori large $p$ and $q$ is to start with its definition \eqref{figuratebinomial} in terms of binomial coefficients, approximate the factorials using the Stirling formula, consider the sum as a Riemann series converging to an integral over $t=\f  n p$ running in $[0,1]$ and computing its saddle point approximation at large $p$. Keeping $q\gg 1$ fixed and defining the ratio $\lambda=\f qp$, this leads to:
\be
T(p,q)\underset{p\gg1}\sim
\f e{2\pi}p\int_{0}^1\rd t\,
\sqrt{\f{(\lambda+t)(1-t)}{t(\lambda+t-1)}}\,e^{p\Phi_{\lambda}[t]}
\quad\textrm{with}\,\,
\Phi_{\lambda}[t]
=
(\lambda+t)\ln(\lambda+t)-t\ln t-(1-t)\ln(1-t)+(\lambda+t-1)\ln(\lambda+t-1)
\,.
\nn
\ee
The stationary point equation $\pp_{t}\Phi=0$ is a quadratic equation in $t$ and the dominant contribution is its positive root $t_{+}$ for large $p$, i.e. when $\lambda\rightarrow 0$, which provides an asymptotic power law approximation for the figurate number:
\be
\pp_{t}\Phi=0\Leftrightarrow t=t_{\pm}=\f{1-\lambda\pm\sqrt{1+\lambda^2}}2
\,,\quad
t_{+}\underset{\lambda\rightarrow 0^+}\sim
1-\f\lambda 2
\,,\quad
\Phi\big{|}_{t_{+}}\underset{\lambda\rightarrow 0^+}\sim
-\lambda\ln\f\lambda 2
\,,\quad
\pp^2\Phi\big{|}_{t_{+}}\underset{\lambda\rightarrow 0^+}\sim
-\f4\lambda
\,,\quad
T(p,q)\underset{p\gg1}\propto \left(\f{2p}q\right)^q
\,.\nn
\ee
}

This is a priori a new relation satisfied by the 6j-symbol describing its behavior under the exchange of opposite edges of the tetrahedron. It looks in spirit similar to the duality formulas derived in \cite{Freidel:2006qv}, but those apply to the square of the q-deformed 6j-symbols. It would nevertheless be very interesting to bridge between the two approaches: check if it is possible to derive a duality relation of the squared 6j-symbols from the identity above and further see if it can be extended to the q-deformed case by a suitable generalization of the figurate numbers.

\section{Fisher zeros on the Tetrahedron} \label{sec:FisherTet}

We call \emph{critical couplings}\footnotemark{} a set of (a priori complex) couplings $\{Y_e\}$ which annihilate the loop polynomial $P_{T}$ an thus the Ising partition function on the tetrahedron. We  denote them $Y_1^c,.., Y_6^c$ and refer to them as the \emph{Fisher zeros} on the tetrahedron.
\footnotetext{Working on a finite graph as we do in the present work, the Ising partition function is irremediably finite. Nevertheless, correlations (between spins), once properly normalized with an inverse factor of the partition function, can nevertheless diverge: when the partition function $\cI_{\Gamma}[\{y_{e}\}]$ vanishes. This does not happen for real couplings $y_{e}$, since in that case the Boltzmannn weights of the Ising are real and positive so that the Ising partition function is necessarily a positive real number. So the zeros of $\cI_{\Gamma}[\{y_{e}\}]$, called the Fisher zeros, are complex couplings, at which correlations might diverge. By a slight abuse of language, we refer to them as critical couplings, although they do not any a priori link with a scale invariance or continuum limit. Nevertheless, we will see later in the next section \ref{sec:poles} that those critical couplings do correspond to scale invariant saddle points for the geometry of spin network evaluations, so they might have an interpretation in terms of continuum limit at the end of the day. Finally, when we take the limit of an infinite regular graph $\Gamma$, we recover the standard notion of critical couplings, for which the Ising partition function will diverge, signaling a fixed point of the coarse-graining flow of the model.
}

An obvious way to find the Fisher zeros consists in expressing one of the couplings $Y_e$ in term of the others, for instance,
\begin{equation}
Y_6 = -\frac{1 + Y_1 Y_3 Y_5 + Y_2 Y_3 Y_4 + Y_1 Y_2 Y_4 Y_5}{Y_1 Y_2 + Y_4 Y_5 + Y_3 (Y_1 Y_4 + Y_2 Y_5)},
\end{equation}
as long as the denominator does not vanish. When the denominator vanishes, the equation $P[Y_e]=0$ simplifies and one can proceed similarly, and so on. This parametrization is not really convenient, as it breaks the symmetry between the variables.

A standard method to parametrize algebraic varieties consists in looking at the intersection of the variety $P[Y_e]=0$ with the lines through the origin. One sets $Y_e = t_e Y$ with a global scale factor $Y$. The equation $P[Y_e]=0$ determines $Y$ as a function of $t_1, \dotsc, t_6$,
\begin{equation}
1 + (t_1 t_2 t_6 + t_1 t_3 t_5 + t_2 t_3 t_4 + t_4 t_5 t_6) Y^3 + (t_1 t_2 t_4 t_5 + t_1 t_3 t_4 t_6 + t_2 t_3 t_5 t_6) Y^4 = 0
\,.
\end{equation}
This gives a symmetric parametrization $Y_e = t_e Y(t_1, \dotsc, t_6)$ of the critical couplings, which however does not seem to have any relation to geometry.

Here we propose a classification of the Fisher zeros, by showing that there is a subset of them  admits a geometrical description in terms of the tetrahedron geometry. To this purpose, we describe below in the next section \ref{sec:PreGeometrical} a more convenient ``pre-geometric'' parametrization of the Fisher zeros. It generalizes the method used in Section \ref{sec:TriangleCycleVariables} to find all complex roots of the polynomials $P_{\Theta}$ and $P_\Delta$. Then a real section of those zeros will admit a parametrization in terms of the triangle and dihedral angles of tetrahedra as explained later in section \ref{sec:geometricfisher}.

As in the case of $P_\Theta$, the general formula for the roots of the polynomial, i.e. our pre-geometric parametrization, is found to be a continuation of the true geometric one to complex edge lengths.

\subsection{Pre-Geometrical Parametrization}
\label{sec:PreGeometrical}

The loop polynomial for the tetrahedron graph $T$ is defined as a sum over all cycles on the graph:
\be
P_{T}[Y_{1,2,3,4,5,6}]=\sum_{C\subset T} L_{C}
\quad\textrm{with}\quad
L_{C}=\prod_{e\in C} Y_{e}\,.
\ee
If we focus on the cycle variables $L_{C}$ and single out the trivial cycle $L_{\varnothing}=1$ for $C=\varnothing$, the loop polynomial reads simply $P_{T}=1+\sum_{C\ne \varnothing}L_{C}$ and identifying the roots of $P_{T}$ in terms of the $L_{C}$'s becomes straightforward. So the goal becomes to express the original edge variables $Y_{e}$ in terms of the cycle variables $L_{C}$.

More precisely, the tetrahedron has six edges for seven non-trivial cycles. There are four 3-cycles and three 4-cycles.
The 3-cycles correspond to the triangles of the tetrahedron graph, as one sees on fig.\ref{fig:6jGraph}, and form a basis of the space of cycles. Considering an edge $e$, let us call $t_1(e), t_2(e)$ the two triangles containing $e$, and $\gamma(e)$ the unique 4-cycle which does not contain $e$. Then, the relation between these three cycles allows to recover the edge coupling $Y_{e}$:
\begin{equation}
L_{t_1(e)} L_{t_2(e)} = Y_e^2 L_{\gamma(e)}.
\end{equation}
For instance, $L_{126} L_{135} = Y_1^2 L_{2356}$. This gives $Y_e^2$ in terms of the cycle variables.

Since the tetrahedron has seven non-trivial cycles for only six edges, there is one more cycle  than edges and there must exist a constraint between the cycle variables. It is quite elementary, the product over the 3-cycles equals the product over the 4-cycles:
\begin{equation}
 \label{CycleConstraint}
L_{126} L_{135} L_{234} L_{456} = L_{1245} L_{1346} L_{2356}
\,.
\end{equation}
As a consequence, we can use the cycle variables $L_{C}$ to trivialize the equation $P_{T}=0$, at the cost of having the constraint \eqref{CycleConstraint} instead.
So one solves $P_{T}=0$ by expressing the $L_{C}$'s in terms of \emph{homogeneous cycle variables} $M_{C}$ such that:
\begin{equation}
\label{LoopParametrization}
L_C = -\frac{M_C}{\sum_{C'} M_{C'}}\quad \text{with the constraint}\quad
M_{126} M_{135} M_{234} M_{456} = -\Bigl(\sum_C M_C\Bigr) M_{1245} M_{1346} M_{2356}
\,.
\end{equation}
The critical couplings $Y_e^c$ are then obtained from the $M_{C}$'s as:
\begin{equation}
\bigl(Y_e^c\bigr)^2 = \frac{M_{t_1(e)} M_{t_2(e)}}{M_{\gamma(e)} \sum_C M_C}.
\end{equation}

This is not yet the pre-geometrical parametrization that we are looking for. We can go further and simplify the constraint on the homogeneous variables. To this purpose, we introduce \emph{complex length} variables $l_1, \dotsc, l_6$ plus one additional variable $n$ defined by the following relations,
\begin{equation} \label{LengthsChangeofVar}
l_e = \sum_{C \ni e} M_C \qquad \text{and}\qquad n = \sum_C M_C.
\end{equation}
Each $l_e$ consists in the sum of two 3-cycles and two 4-cycles, for instance, $l_1 = M_{126} + M_{135} + M_{1245} + M_{1346}$.
The structure of the cycles of the \emph{dual} tetrahedron $T^*$ comes into play when inverting those relations. If $t$ is 3-cycle, i.e. a triangle of the tetrahedron $T$, we call $t^*$ the triangle in the dual graph consisting in the three opposite edges, which are all attached to the vertex $v$ opposite to $t$. For instance, the triangle $t=(456)$ gives the opposite triangle $t^*=(123)$ in the dual tetrahedron.
%
One can then write the variables $M_{t}$ for the 3-cycles in terms of the (complex) \emph{semiperimeter} $s_{t^*}$ around the opposite triangle $t^*$,
\begin{equation}
M_{t} = n - s_{t^*}
\,,
\qquad
s_{t^*} = \frac12 \sum_{e\in t^*} l_e
\,.
\end{equation}
For instance, $ M_{456}=n-\f12(l_1 + l_2 + l_3)$. As for the 4-cycles $\gamma$, we write similarly\footnotemark:
\begin{equation}
M_\gamma = s_\gamma - n,\qquad \text{with}\qquad s_\gamma = \frac12\sum_{e\in \gamma} l_e
\,.
\end{equation}
which gives for instance $M_{1245}=\f12(l_1 + l_2 + l_4 + l_5) - n$.
\footnotetext{
Notice that instead of labeling $s_{t^*}$ by the triangle $t^*$, we could have used its dual vertex $v\in T$. It seems  however more natural to use $t^*$ since this allows to unify the definition of the variables $s_{t^*}$ and $s_\gamma$ (recall that $\gamma$ is the same 4-cycle in both $T$ and $T^*$) as  complex semi-perimeters of cycles, $s_{C^*} = \tfrac12 \sum_{e\in C^*} l_e$, where $C^*$ is a cycle in $\Gamma^*$.
}
Since the six complex lengths together with the variable $n$ form a set of seven variables, the constraint \eqref{CycleConstraint} still applies. It now becomes:
\begin{equation}
(n - s_{123}) (n - s_{156}) (n - s_{345}) (n - s_{246}) = -n (s_{1245} - n) (s_{1346} - n) (s_{2356} - n).
\end{equation}
While this equation looks like a quartic equation in $n$, it is immediate to see that the coefficients of $n^4$ and $n^3$ vanish. It reduces to
\begin{equation}
\label{SaddleEqRacah}
a n^2 + b n + c = 0,
\end{equation}
with the coefficients $a$, $b$ and $c$ given in terms of the complex lengths $l_{e}$:
\begin{equation}
a = \frac{1}{2}(l_1 l_4 + l_2 l_5 + l_3 l_6),\quad
b = -\frac14 (l_1 l_4 + l_2 l_5 + l_3 l_6) \Bigl(\sum_e l_e\Bigr) - \frac14\sum_{t^*} \prod_{e\in t^*} l_e, \quad
c = \prod_{t^*} s_{t^*}
\,.
\end{equation}

Putting everything together, we start with six complex lengths $l_{e}$.
The variable $n$ takes two possible values, determined by the quadratic equation \eqref{SaddleEqRacah} as a function of the six complex lengths.
And we arrive at the pre-geometrical parametrization for the roots of the loop polynomial $P_{T}$, i.e. the Fisher zeros on the tetrahedron:
\begin{equation}
\label{pregeom}
\bigl(Y_e^c\bigr)^2 = \frac{(n - s_{t_1^*(e)}) (n - s_{t_2^*(e)})}{n (s_{\gamma(e)} - n)}
\end{equation}
where $t_1^*(e), t_2^*(e)$ are the two triangles in the dual tetrahedron $T^*$ which do not contain the edge $e$ (for instance, the edge $e=1$  belongs to the two triangles $t_{1}=(126)$ and $t_{2}=(135)$, which defines the two opposite triangles $t_{1}^*=(345)$ and $t_{2}^*=(246)$ in the dual tetrahedron), and $\gamma(e)$ is the 4-cycle which also avoids $e$.  
Let us underline that in general the variables $s_{t^*}, s_\gamma$ and $n$ are all complex numbers.

\subsection{Geometrical Fisher zeros}
\label{sec:geometricfisher}

We now show that the restriction of the above ansatz to a specific domain of the complex lengths leads to solutions of $P_{T}=0$ parametrized by the geometry of the dual tetrahedron $T^*$.
We underline that since we are working with the spin network evaluations on the tetrahedron graph $T$, it is natural to interpret them in terms of the geometry of the dual graph $T^*$. Indeed, the 3j symbol, or intertwiner, living at a vertex of the original graph $T$ is usually interpreted as a quantum geometry for the dual triangle in $T^*$. Then the 6j-symbol defined on the tetrahedron graph $T$ is interpreted in terms of the volume and Regge action for the geometry of the dual tetrahedron $T^*$.

More precisely, we restrict the complex edge lengths to real positive lengths $l_1, \dotsc, l_6$ satisfying the triangle inequalities for the triangles of $T^*$. We further require that the discriminant of the quadratic equation \eqref{SaddleEqRacah} is real and negative. This discriminant reads:
\beq
b^2 - 4 a c
&= -\frac{1}{16} &
\Big{[}l_1^2 l_4^2 (l_2^2 + l_5^2 + l_3^2 + l_6^2 - l_1^2 - l_4^2) + l_2^2 l_5^2 (l_1^2 + l_4^2 + l_3^2 + l_6^2 - l_2^2 - l_5^2) + l_3^2 l_6^2 (l_1^2 + l_4^2 + l_2^2 + l_5^2 - l_3^2 - l_6^2)
\nn
\\
&&- (l_1^2 l_2^2 l_3^2 + l_3^2 l_4^2 l_5^2 + l_1^2 l_5^2 l_6^2 + l_2^2 l_6^2 l_4^2)\Big{]}\,,
\label{Vsquare}
\eeq
where we recognize the squared volume $V^2$ of the tetrahedron with lengths $l_1, \dotsc, l_6$, as given by the Cayley-Menger determinant,
\begin{equation}
b^2 - 4 a c = -9 V^2.
\end{equation}
Requiring that the discriminant be negative amounts to requiring that the squared volume $V^2$ be positive. This is actually not automatically ensured by assuming that the edge lengths satisfy the triangular inequalities. In fact, choosing edge lengths satisfying the triangular inequalities does not always ensure the existence of a corresponding tetrahedron in 3d Euclidean space. This is only true when $V^2>0$. When $V^2<0$, it is actually possible to build a tetrahedron in 3d Lorentzian space from those edge lengths \cite{Barrett:1993db}. This corresponds to the exponentially decreasing branch of the asymptotics of the 6j-symbols. We postpone to future analysis  the case $V^2<0$, which could provide a geometrical interpretation of the Fisher zeros in terms of such Lorentzian tetrahedra.

\medskip

As we assume the squared volume $V^2$ to be positive and thus the discriminant $(b^2-4ac)$ to be negative, the solutions to \eqref{SaddleEqRacah} for the parameter $n$ are
\begin{equation}
n_\pm = \frac{-b \pm 3i V}{2a}.
\end{equation}
The fact that $a, b, V$ are all real allows for a nice geometrical form of the critical couplings $Y_e^c$. First, let us plug the solution $n_\pm$ in the formula for the critical couplings:
\begin{equation}
\bigl(Y_e^c\bigr)^2 = \frac{(-b - 2a s_{t_1^*(e)} \pm 3iV)(-b - 2a s_{t_2^*(e)} \pm 3iV)}{(-b \pm 3iV)(2a s_{\gamma(e)} + b \mp 3iV)}
\,.
\end{equation}
Here, $t_1^*(e), t_2^*(e)$ are the two triangles in the dual tetrahedron $T^*$ which do not contain the edge $e$, so $(345)$ and $(246)$ for the edge $e=1$.
Notice that switching signs $n_{+}\leftrightarrow n_{-}$ amounts to taking the complex conjugate of the critical couplings $Y_{e}^c\leftrightarrow \bar{Y}_{e}^c$.

\begin{figure}[h!]

\centering

\begin{tikzpicture}[scale=0.75]

\coordinate(A) at (0,0);
\coordinate(B) at (2.5,3.5);
\coordinate(C) at (3.5,-1);
\coordinate(D) at (5.5,.5);

\draw (A)--node[pos=0.7,left]{$i^1(e)$}(B) ;
\draw (A)--node[pos=0.5,below]{$j^1(e)$}(C) ;
\draw (D)--node[pos=0.65,right]{$i^2(e)$}(B) ;
\draw (D)--node[pos=0.6,right]{$j^2(e)$}(C) ;

\pic [draw, "$\phi_{\Delta^1(e)}$", angle eccentricity=2.5,angle radius=10] {angle = C--A--B};
\pic [draw, "$\phi_{\Delta^2(e)}$", angle eccentricity=2.3,angle radius=10] {angle = B--D--C};

\draw (B)-- node[pos=0.4,left]{$\Delta^1(e)$}node[pos=0.4,right]{$\Delta^2(e)$}node[pos=0.7,left]{$e$} (C);


\end{tikzpicture}

\caption{
For an arbitrary edge $e$ in the dual tetrahedron $T^*$, we denote the two triangles sharing it $\Delta^a(e)$,  its adjacent edges $i^a(e), j^a(e)$ and opposite angles $\phi_{\Delta^a(e)}$ in $T^*$, for $a=1,2$.
} 
\label{fig:AdjacentTriangles}

\end{figure}
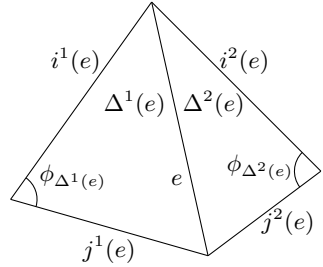

All the quantities on the right hand side are explicit functions of the six lengths, so we can extract the norm and phase of $Y_e^c$. %
In practice, we start by performing the explicit calculations for one edge, say $e=1$ for simplicity's sake, then extend the result to an arbitrary edge using the symmetries of the tetrahedron. We compute the norm of $Y_{1}^c$:
\begin{equation*}
\big|Y_1^c\big|^2 = \sqrt{\frac{(l_1 - l_2 + l_3)(l_1 + l_2 - l_3)\ (l_1 - l_5 + l_6)(l_1 + l_5 - l_6)}{(-l_1 + l_2 + l_3)(l_1 + l_2 + l_3)(-l_1 + l_5 + l_6)(l_1 + l_5 + l_6)}}.
\end{equation*}
For any edge $e$, we call $\Delta^1(e)$ and $\Delta^2(e)$ the two triangles in the dual tetrahedron $T^*$ sharing the edge $e$, and we denote $i^1(e), j^1(e)$ the two other edges in $\Delta^1(e)$, and similarly $i^2(e), j^2(e)$ in $\Delta^2(e)$. Then,
\begin{equation}
\begin{aligned}
|\bigl(Y_e^c\bigr)^2| &=
 \sqrt{\frac{(l_e - l_{i^1(e)} + l_{j^1(e)})(l_e + l_{i^1(e)} - l_{j^1(e)})}{(-l_e + l_{i^1(e)} + l_{j^1(e)})(l_e + l_{i^1(e)} + l_{j^1(e)})}}
  \sqrt{\frac{(l_e - l_{i^2(e)} + l_{j^2(e)})(l_e + l_{i^2(e)} - l_{j^2(e)})}{(-l_e + l_{i^2(e)} + l_{j^2(e)})(l_e + l_{i^2(e)} + l_{j^2(e)})}}
 \\
&=
\sqrt{\frac{(s_{\Delta^1(e)} - l_{i^1(e)})(s_{\Delta^1(e)} - l_{j^1(e)}) }{s_{\Delta^1(e)} (s_{\Delta^1(e)} - l_e)}}
\sqrt{\frac{ (s_{\Delta^2(e)} - l_{i^2(e)})(s_{\Delta^2(e)} - l_{j^2(e)})}{s_{\Delta^2(e)} (s_{\Delta^2(e)} - l_e)}}
\,,
\end{aligned}
\end{equation}
with the $s_{\Delta}$'s still standing for the half-perimeters of the corresponding triangle (in $T^*$).
That can easily be further simplified and expressed in terms of the 2d angles of the triangles, as we can recognize the same formulas as in Section \ref{sec:TriangleCycleVariables} for the tangents of the half-angles of $\Delta^1(e)$ and $\Delta^2(e)$.
Indeed, in a triangle with edges $(ijk)$, the cosine of the angle between the edges $i$ and $j$ is given by
\begin{equation}
\cos \phi_{ij} = \frac{l_i^2 + l_j^2 - l_k^2}{2\,l_i l_j},\qquad
\text{or equivalently} \qquad
1-\cos\phi_{ij} = \frac{(-l_i + l_j + l_k)(l_i - l_j + l_k)}{2\,l_i l_j},
\end{equation}
while the sine of that angle can be recovered by the Heron formula for the triangle area:
\begin{equation} \label{SinPhi}
\sin \phi_{ij} = \frac{\sqrt{(l_i + l_j + l_k)(l_i + l_j - l_k)(l_i - l_j + l_k)(-l_i + l_j - l_k)}}{2\,l_i l_j}
\,,
\end{equation}
so that we find for the tangent of the half-angle:
\begin{equation}
\tan \frac{\phi_{ij}}{2} = \frac{1-\cos\phi_{ij}}{\sin \phi_{ij}} = \sqrt{\frac{(-l_i + l_j + l_k)(l_i - l_j + l_k)}{(l_i + l_j - l_k)(l_i + l_j + l_k)}} = \sqrt{\frac{(s_{ijk} - l_i)(s_{ijk} - l_j)}{s_{ijk} (s_{ijk} - l_k)}},
\end{equation}
where $s_{ijk} = \tfrac12(l_i + l_j + l_k)$ is the semi-perimeter of the triangle.
This shows that the norm of the critical coupling $Y_e^c$ is determined by the two 2d angles opposite to the edge $e$ in the tetrahedron $T^*$, i.e. the angles $\phi_{\Delta^1(e)}$ opposite to $e$ in $\Delta^1(e)$ and $\phi_{\Delta^2(e)}$ opposite to $e$ in $\Delta^2(e)$, as shown in Figure \ref{fig:AdjacentTriangles},
\begin{equation}
\big|Y_e^c\big|^2 =
\tan \frac{\phi_{\Delta^1(e)}}{2}\ \tan \frac{\phi_{\Delta^2(e)}}{2}
\,.
\end{equation}

To extract the phase of the critical couplings, we start once again with the edge $e =1$, and evaluate the ratio of its real part to its norm,
\begin{equation} \label{RealPartYc}
\frac{\Re \Bigl(\bigl(Y_1^c\bigr)^2\Bigr)}{\big|Y_1^c\big|^2} = \frac{2 l_1^2 l_4^2 + l_1^2 (l_1^2 - l_2^2 - l_3^2 - l_5^2 - l_6^2) + (l_2^2 - l_3^2)(l_6^2 - l_5^2)}{16\,A_{123}\,A_{156}}.
\end{equation}
Here $A_{ijk}$ is the area of the triangle with edges $(ijk)$ given in terms of the edge lengths by the Heron formula, as in \eqref{Heron}.
The right hand side of \eqref{RealPartYc} is recognized\footnotemark{} as the cosine of the \emph{external dihedral angle} $\theta_1$ hinged at $e=1$.
\footnotetext{
To compute the dihedral angle, we can express it in terms of the 2d angles $\cos \theta_i = \tfrac{\cos \phi_{ij} \cos \phi_{ik} - \cos \phi_{jk}}{\sin \phi_{ij} \sin \phi_{ik}}$ if the edges $i,j,k$ meet at a vertex of the tetrahedron $T^*$. Then using the explicit formulae for the 2d angles $\cos \phi_{ee'}$ and $\sin \phi_{ee'}$, one obtains an explicit formula for $\cos \theta_i$ in terms of the edge lengths.
}
In general, we get:
\begin{equation}
\frac{\Re \Bigl(\bigl(Y_e^c\bigr)^2\Bigr)}{\big|Y_e^c\big|^2} = \cos \theta_e.
\end{equation}
Moreover, since swapping $n_+ \leftrightarrow n_-$ turns the couplings $Y_e^c$ into their complex conjugate, we obtain the two possible signs for their imaginary parts,
\begin{equation}
\frac{\Im \Bigl(\bigl(Y_e^c\bigr)^2\Bigr)}{\big|Y_e^c\big|^2} = \pm \sin \theta_e.
\end{equation}
This gives  a fully geometrical expression for the critical couplings parametrized by the six lengths of an arbitrary Euclidean tetrahedron:
\begin{equation} \label{GeometricSol}
\forall e=1, \dotsc, 6 \qquad Y_e^c = \exp\Bigl(\epsilon \frac{i}{2} \theta_e\Bigr) \sqrt{\tan \frac{\phi_{\Delta^1(e)}}{2}\ \tan \frac{\phi_{\Delta^2(e)}}{2}}
\,, 
\end{equation}
with an arbitrary overall sign $\epsilon=\pm$. Since this expression only depends on the angles of the tetrahedron, these critical couplings only depend on the geometry of the tetrahedron up to a global scale factor and thus depend on five real parameters.

Let us start with a couple of technical comments. First, the geometric formula \eqref{GeometricSol} for the critical couplings describes the restriction of the general formula \eqref{pregeom} to real edge lengths satisfying triangular inequalities and giving a positive squared volume. In some sense, this defines a 5-dimensional real section of the full 10-dimensional sets of complex solutions to $P_{T}[Y_{1,..,6}]=0$. It would be interesting to see if one provides arbitrary complex solutions with a similar geometric interpretation, similarly to the Cevian parametrization of the complex critical couplings for the $\Theta$ graph in terms of triangles and Cevians presented earlier in section \ref{cevian}.

Moreover, the duality map $\cD$ sends roots of the loop polynomial $P_{T}$ to roots of its dual $P_{T^*}$. However it is not clear how the pre-geometric parametrization \eqref{pregeom} behaves under this duality map. Moreover, numerical checks show that the duals of geometric Fisher zeros on $T$ built from \eqref{GeometricSol}  are not geometric Fisher zeros on $T^*$: in some sense, the duality map $\cD$ send does not ``real'' Fisher zeros to ``real'' Fisher zeros on the dual graph. This effect was already shown on the $\Theta$ graph and its dual triangular graph. It would nevertheless be interesting to see if it is possible to derive the pre-geometric parametrization of the dual critical couplings or at least provide the dual of the geometric critical couplings with their own geometric interpretation.

Now let us make broader comments.
\begin{itemize}
\item Remarkably, the formula for the norm of the critical couplings $|Y_e^c|$ is known to be significant for the Ising model on isoradial graphs \cite{BoutillierDeTiliereSurvey}. In that case the opposite angles for each edge $e$ are equal $\phi_{\Delta^1(e)} = \phi_{\Delta^2(e)} \equiv \phi_e$ and $|Y_e^c| = \tan (\phi_e/2)$. For those values of the couplings, $Y_e = \tan (\phi_e/2)$,  Baxter found that the 2D Ising model on isoradial graphs becomes critical in the thermodynamical limit. Notice however that we have derived our formula here  only for the tetrahedron graph and can not consider a thermodynamical limit. Nevertheless, we have already derived the same formula for the modulus of the critical couplings $|Y_e^c|$ for an arbitrary graph in terms of the tangent of the opposite half-angles in a previous work \cite{Bonzom:2015ova}. In light of the present work, our new result for the tetrahedron graph hints to a general geometric formula for the critical couplings of the Ising model in terms of 2D and 3D angles.

\item Our result deepens the relevance of geometry for the Ising model. The critical couplings on isoradial graphs are real and depend solely on the 2D geometry of the graph. In our case, the phase $e^{\pm i\theta_e/2}$ adds an extra dependence on the \emph{extrinsic geometry}, i.e. the way the graph is embedded into 3D space. For a finite planar graph, its embedding into 3D space will necessarily produce non-trivial  dihedral angles, so we expect generically non-trivial phases and complex Fisher zeros. In the thermodynamic limit, when the graph becomes infinite, it is possible to make it flat and set all the dihedral angles to 0, thus obtaining real critical couplings. One can nonetheless choose a non-trivial embeddings into 3D space, even in the thermodynamical limit, and then obtain complex critical couplings according to the proposed formula above.

This nevertheless requires to generalize the work done here for the tetrahedron graph to arbitrary graphs. The derivation presented in this section requires a detailed understanding and knowledge of the combinatorics of the cycles on the tetrahedron graph and of the tetrahedron geometry. Producing a similar proof for an arbitrary graph seems quite challenging. In a previous work  \cite{Bonzom:2015ova}, we proposed an alternative method exploiting the relation between the Ising partition function and the  generating function for spin network evaluations, as given by \eqref{westbury}: one identifies the Ising critical couplings  as the poles of the generating function for spin network evaluations, which can be extracted through a saddle point analysis. We test this method for the case of the tetrahedron graph below in section \ref{sec:poles} and show that we recover exactly the same geometric formula \eqref{GeometricSol} for the critical couplings.

\end{itemize}

\subsection{A numerical check: the equilateral tetrahedron}

Let us apply the geometric formula for the Fisher zeros to the simplest case of an equilateral tetrahedron. This should give the homogeneous Fisher zeros, that is the roots of the loop polynomial $P_{T}[Y_{1,..,6}]=0$ when all six edge couplings are equal, $Y_{1}=..=Y_{6}$.

The geometry of the equilateral tetrahedron is well-known. We need the 2D angles, within each triangle:
\be
\phi=\f\pi3,
\quad
\cos\phi=\f12,
\quad
\sin\phi=\f{\sqrt{3}}2,
\quad
\tan\f\phi2=\f1{\sqrt{3}}
\,,
\ee
and the 3D (external) dihedral angles:
\be
\cos\theta
\,=\,
-\f{\cos\phi -\cos^2\phi}{\sin^2\phi}
\,=\,
-\f13\,
\ee
The 2D and 3D angles are combined to get the modulus and phase of the corresponding geometric roots:
\be
Y_{\pm}=\f1{\sqrt{3}}e^{\pm i\f\theta2}
=\f13\,(1\pm i\sqrt{2}),
\quad
\cD(Y_{\pm})=Y_{\mp},
\quad
P(Y_{\pm})=0\,,
\ee
which are complex roots of the homogeneous loop-polynomial $P_{T}[Y]=1+4Y^3+3Y^4=(Y+1)^2(3Y^2-2Y+1)$ when the edge couplings are take all equal $Y_{1}=..=Y_{6}=Y$. The other root is the trivial solution $Y^0=-1$ corresponding to the degenerate case when the Ising couplings $y_{e}$ are sent to $-\infty$ thus forcing all the Ising spins to be synchronized.

Let us nevertheless point out that the modulus of the homogeneous tetrahedron Fisher zeros $|Y_{\pm}|=1/\sqrt{3}$ is exactly the critical coupling for the 2D Ising model on the regular honeycomb lattice. This is understandable since this norm is built from the 2D geometry of the equilateral triangles. The phases of the critical couplings, here $e^{\pm i\f\theta2}$, represent the dihedral angles between those triangles, thus how the lattice is folded within the 3D space. In a thermodynamical limit where the lattice is embedded as totally flat in the 3D space, the phase can be trivial and we are left with a real critical coupling entirely determined by the 2D geometry, so $Y^c=1/\sqrt{3}$ for the regular honeycomb lattice.

\section{Fisher zeros as Poles: Large Spin Asymptotics \& Saddle Points}
\label{sec:poles}

In this section, we show how to recover the geometric critical couplings \eqref{GeometricSol} derived in the previous section as poles of the generating function for spin network evaluations. This underlines a new method to derive Fisher zeros for the 2D Ising model on arbitrary graphs.

The dependence of the geometric solutions $Y_e^c$ on the extrinsic geometry is natural from the point of view of spin networks. Indeed, the Hamiltonian constraint (Einstein equation written in the Hamiltonian formalism) describes how the intrinsic geometry relates to the extrinsic curvature. In 3D loop quantum gravity and Regge calculus for flat space-time, it means that all dihedral angles can be calculated from the 2D geometry using the standard formulae of 3D Euclidean (flat) geometry.
As well-known, this is reflected at the quantum level in the asymptotics of physical states and transition amplitudes between spin networks, e.g. \cite{Bonzom:2011hm,Bonzom:2011nv}. Both for physical states and transition amplitudes defined by the Ponzano-Regge model, the basic building block is  Wigner's 6j-symbol, whose asymptotics in the oscillatory regime ($V^2>0$) is given by \cite{PR1968}\footnotemark:
\be \label{PRFormula}
\sixj{j_{1}}{j_{2}}{j_{3}}{j_{4}}{j_{5}}{j_{6}}
\sim
\f1{\sqrt{12 \pi V}}\,\cos \left(S_{R}(\{j_e\})+\f\pi4\right)
\qquad\text{with}\quad
S_{R}(\{j_e\}) = \sum_{e=1}^6 (j_{e}+\f12)\theta_{e}\,,
\nn
\ee
$S_R$ is the Regge action for the tetrahedron with edge lengths $l_{e}=(j_{e}+\f12)$ and the $\theta_{e}$ are the tetrahedron dihedral angles as functions of the edge lengths.
\footnotetext{
 The interested reader can also check \cite{Schulten:1971yv} for a proof based on the recursion relations satisfied by the 6j-symbol, \cite{Roberts:1998zka} for a proof based on geometric quantization, \cite{Barrett:2002ur,Freidel:2002mj} for proofs based on group integrals over the $\SU(2)$ Lie group and \cite{Gurau:2008yh,Dupuis:2009sz} for a proof directly from the combinatorial expression of the 6j-symbols in terms of factorials.
 }

In the previous sections, we have found the critical couplings $Y_e^c$ via a direct analysis of the equation $P_{T}=0$. We propose that the critical couplings $Y_e^c$, especially the geometric formula, could also be found using spin network evaluations.
Actually, the zeros of the loop polynomial $P_{T}$ correspond to the poles of  $1/P_{T}^2$, which is the generating function of 6j-symbols. Therefore a saddle point analysis of the summand of the generating function should reveal the loci of the poles.

Moreover, the equation \eqref{SaddleEqRacah} determining the pre-geometric parametrization of Fisher zeros on the tetrahedron graph turns out to be exactly the saddle point equation for the asymptotic evaluation of Racah formula for the 6j-symbol, as used in \cite{Gurau:2008yh} to re-derive the 6j-asymptotics \eqref{PRFormula}. More details  can be found in appendix \ref{loopexpansion}.

\bigskip

So let us look into the saddle equation for the generating function of the spin network evaluations on the tetrahedron graph:
\be
\frac{1}{P_{T}[\{Y_e\}]^2}
\,=\,
\cZ_{T}[\{Y_e\}]
\,=\,
\sum_{\{j_{e\}}} \prod_{v}\Delta_v(j_e)\,
\sixj{j_{1}}{j_{2}}{j_{3}}{j_{4}}{j_{5}}{j_{6}} \,\prod_{e}Y_{e}^{2j_{e}}
\,.
\ee
We would like to analyze the behavior of this series in the $j_{e}$'s. We assume that we can work in the large spin asymptotics. Actually, both the asymptotics of the factorials in the weights $\Delta_v(j_e)$ and of the 6j-symbols quickly approximate in a pretty correct and precise fashion their behavior, as soon as the $j$'s are larger than 10 (see e.g. \cite{Livine:2006ab} for a numerical analysis of the precision of the asymptotics of the 6j-symbols).

We start with the factorial weights, which  can be approximated using Stirling's formula:
\be
\prod_{v}\Delta_v(j_e)=\sqrt{\prod_{v}\f{(J_{v}+1)!}{\prod_{e\ni v}(J_{v}-2j_{e})!}}
\underset{j_{e\gg 1}}\sim 
e^{\Phi[\{j_{e}\}]}\,,
\quad\textrm{with}\quad
\Phi
=\sum_{v} \left[J_{v}\ln J_{v} -\sum_{e\ni v} (J_{v}-2j_{e})\ln(J_{v}-2j_{e})
\right]
\,.
\ee
Then the asymptotics of the 6j-symbols, for a homogeneous large rescaling of the spins\footnotemark, behave as:
\footnotetext{We rescale the six spins by the same factor $j_i = \lambda {j}^o_i$. The homogeneous large spin limit is keeping the initial spins ${j}^o_i$ fixed and sending the overall factor $\lambda$ to $\infty$.}
\be
\sixj{j_{1}}{j_{2}}{j_{3}}{j_{4}}{j_{5}}{j_{6}}
\sim
\f1{\sqrt{12\pi V[\{j_{e}\}]}}\,\cos \left(S_{R}[\{j_{e}\}]+\f\pi4\right)\,,
\quad\textrm{with}\quad
S_{R}[\{j_{e}\}]=\sum_{e} (j_{e}+\f12)\theta_{e}[\{j_{e'}\}]\,,
\ee
where $V[\{j_{e}\}]$ and $S_{R}[\{j_{e}\}]$ are respectively the volume and the Regge action for the dual tetrahedron $T^*$ with edge lengths $l_{e}=j_{e}+\f12$. The angle $\theta_{e}$ is the external dihedral angle hinged at the edge $e$ and is considered as a function of the edge lengths.
For this asymptotics to hold, we have assumed that the squared volume $V^2$, computed from the edge lengths (e.g. as given by \eqref{Vsquare}) is positive.
The $+\f12$ is a next-to-leading order correction, it crucially improves the numerical fit of both the phase and modulus of the 6j-symbol asymptotics but we can safely discard it in the present saddle point analysis at leading order.
So the asymptotics of the 6j-symbol is described in terms of the geometry of the dual tetrahedron $T^*$ with edge lengths given by the spins, $l_{e}=j_{e}$.

If we split the cosine into two complex exponentials labeled by a sign $\eps=\pm$,
\be
\cos \left(S_{R}+\f\pi4\right)
=\f12\sum_{\eps=\pm} e^{i\eps\left(S_{R}+\f\pi4\right)}
\,,\qquad
S_{R}\sim\sum_{e}j_{e}\theta_{e}
\,,
\nn
\ee
we can write the generating function in term of an action:
\be
\prod_{v}\Delta_v(j_e)\,
\sixj{j_{1}}{j_{2}}{j_{3}}{j_{4}}{j_{5}}{j_{6}} \,\prod_{e}Y_{e}^{2j_{e}}
\underset{j_{e}\gg 1}\sim
\f1{2\sqrt{12\pi V[\{j_{e}\}]}}\sum_{\eps=\pm}
e^{i\eps\f\pi4}e^{\cS_{\eps}[\{j_{e}\},\{Y_{e}\}]}
\,,
\ee
\be
\cS_{\eps}[\{j_{e}\},\{Y_{e}\}]
=
\Phi[\{j_{e}\}] + \sum_e 2j_e\ln |Y_e| + i\sum_e j_e(\epsilon  \theta_e  -  2 \Theta_e)
\,,
\ee
where we have distinguished the norm from the phase of the edge couplings, $Y_{e} = |Y_{e}| e^{-i\Theta_{e}}$.
The modulus $|Y_{e}|$ only couples to the measure factor $\prod_{v}\Delta_{v}$, while the phase  couples to the Regge action coming from the 6j-symbol.

\begin{figure}[h!]

\begin{subfigure}[t]{.35\textwidth}
\begin{tikzpicture}[scale=2.5]

\coordinate(a) at (0,0);
\coordinate(b) at (1,0.15);

\draw (a)--(b)node[midway,above]{${e}$};

\draw (a)++(0.15,-0.1)node {$v^1(e)$} ;
\draw (b)++(-0.1,-0.13) node {$v^2(e)$} ;

\coordinate(a1) at (-0.6,0.35);
\coordinate(a2) at (-0.6,-0.37);
\coordinate(b1) at (1.6,0.5);
\coordinate(b2) at (1.6,-0.2);

\draw (a)--(a1)node[midway,above]{$i^1(e)$};
\draw (a)--(a2)node[midway,below]{$i^2(e)$};
\draw (b)--(b1)node[midway,above]{$j^1(e)$};
\draw (b)--(b2)node[midway,below]{$j^2(e)$};

\end{tikzpicture}
\caption{For an edge $e$ in the tetrahedron graph $T$, we call $v_{a}(e)$ the two vertices at its extremities and we write $i^a(e)$, $j^a(e)$ for the four  edges to which it is connected.}
\end{subfigure}
\hspace{10mm}
\begin{subfigure}[t]{.45\textwidth}
\begin{tikzpicture}[scale=0.85]

\coordinate(A) at (0,0);
\coordinate(B) at (2.5,3.5);
\coordinate(C) at (3.5,-1);
\coordinate(D) at (5.5,.5);

\draw (A)--node[pos=0.7,left]{$i^1(e)$}(B) ;
\draw (A)--node[pos=0.5,below]{$j^1(e)$}(C) ;
\draw (D)--node[pos=0.65,right]{$i^2(e)$}(B) ;
\draw (D)--node[pos=0.6,right]{$j^2(e)$}(C) ;

\pic [draw, "$\phi_{\Delta^1(e)}$", angle eccentricity=2.5,angle radius=10] {angle = C--A--B};
\pic [draw, "$\phi_{\Delta^2(e)}$", angle eccentricity=2.3,angle radius=10] {angle = B--D--C};

\draw (B)-- node[pos=0.7,left]{$e$} (C);

\draw[dashed,in=+120,out=+70,looseness=1.5,decoration={markings,mark=at position 0.3 with {\arrow[scale=1.2,>=angle 60]{>}}},postaction={decorate}] (2.2,1.1) to node[above,pos=.25]{$\theta_{e}$} (4,1.3);


\end{tikzpicture}
\caption{An edge $e$ in the tetrahedron graph $T$ is dual to an edge in the dual tetrahedron graph $T^*$, which we also call $e$ by a slight abuse of notations. We call  $\phi_{\Delta^a(e)}$ the triangle angles opposite to the edge $e$ in the two triangles sharing it and $\theta_{e}$ the dihedral angle between the two triangle planes.}

\end{subfigure}

\caption{
Neighborhood of an edge $e$ in the tetrahedron graph $T$ and in the dual tetrahedron $T^*$.
} 
\label{fig:neighbortriangle}

\end{figure}
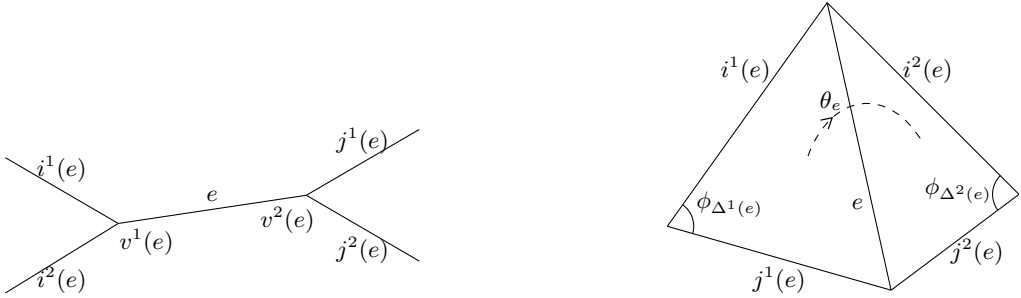

We are interested in the stationary point equation, $\pp_{j_{e}}\cS_{\eps}=0$.  These saddle points should control and dominate the series $\sum_{j_{e}}\exp(\cS_{\eps}[\{j_{e}\},\{Y_{e}\}])$. The essential property of the Regge action is that, although the dihedral angles depend non-trivially on the edge lengths of the tetrahedron, the Schlafl\"i identity, $\sum_{e}j_{e}\rd \theta_{e}=0$, ensures that the derivative of the Regge action $S_{R}$ with respect to an edge length is simply the corresponding dihedral angle:
\be
\pp_{j_{e}}S_{R}[\{j_{e}\}]
=
\pp_{j_{e}}\left[\sum_{e'}j_{e'}\theta_{e'}\right]
=
\theta_{e}
\,.
\ee
Now, the saddle point equation, $\pp_{j_{e}}\cS_{\eps}=0$, gives two equations respectively for the real and imaginary parts of the action.
Clearly, the imaginary part fixes the dihedral angles at a stationary point, in terms of the phase of the couplings, $\Theta_{e}=\eps \theta_{e}/2$, while their modulus of the couplings enter the real part of the stationary point equation,
\be
\pp_{j_{e}} \Re\big{(}\cS_{\eps}[\{j_{e}\}]\big{)}
=
\pp_{j_{e}} \left[\Phi+\sum_{e} 4j_{e}\ln |Y_{e}|\right]
=
0
\,.
\ee
\be
\pp_{j_{e}}
\Re\big{(}\cS_{\eps}[\{j_{e}\}]\big{)}
\,=\,
4\ln |Y_{e}|
+\left[
\ln \f{J_{v^1(e)}(J_{v^1(e)} - 2j_{e})}{(J_{v^1(e)} - 2j_{i^1(e)})(J_{v^1(e)} - 2j_{j^1(e)})}
\right]
+\left[
\ln \f{J_{v^2(e)}(J_{v^2(e)} - 2j_{e})}{(J_{v^2(e)} - 2j_{i^2(e)})(J_{v^2(e)} - 2j_{j^2(e)})}
\right]
\,,
\ee
with the following notations, as illustrated on fig.\ref{fig:neighbortriangle}: $v^1(e)$ and $v^2(e)$ are the two vertices $e$ is incident to, $i^1(e), j^1(e)$ are the two edges meeting $e$ at $v^1(e)$ and similarly for $i^2(e), j^2(e)$ at $v^2(e)$.

At this point, it is better to think with the dual tetrahedron $T^*$ in mind, instead of the original tetrahedron graph $T$, since we are dealing with the geometry of this dual tetrahedron. In this dual context, the two vertices $v^1(e)$ and $v^2(e)$ become the two triangles $\Delta^1(e)$ and $\Delta^2(e)$ which share the edge $e$.
Since the spins $j_{e}$ give the edge lengths $l_{e}$, the sum $J_{v}$ of the spins  around a vertex $v\in T$ gives the perimeter of the triangle in $T^*$ dual to $v$.
%

Thus a stationary point is given by the spins $j_{e}$ being the edge lengths $l_{e}$ of a closed tetrahedron in Euclidean space $\R^3$, related to the modulus of the couplings $|Y_{e}|$ by
\be
|Y_{e}|^2\,=\,
\sqrt{
\f{(s_{\Delta^1(e)}-l_{i^1(e)})(s_{\Delta^1(e)}-l_{j^1(e)})}{s_{\Delta^1(e)}(s_{\Delta^1(e)}-l_{e})}
\,
\f{(s_{\Delta^2(e)}-l_{i^2(e)})(s_{\Delta^2(e)}-l_{j^2(e)})}{s_{\Delta^2(e)}(s_{\Delta^2(e)}-l_{e})}
}
=
\tan\f{\phi_{\Delta^1(e)}}2 \tan\f{\phi_{\Delta^2(e)}}2\,,
\ee
where $\phi_{\Delta^1(e)}$ and $\phi_{\Delta^2(e)}$ are  the two 2D angles opposite to the edge $e$ in the triangles $\Delta^1(e)$ and $\Delta^2(e)$.
Putting the norm and phase together gives:
\be
\label{saddlept}
Y_{e}
=
e^{i\eps \f{\theta_{e}}2}
\,
\sqrt{\tan\f{\phi_{\Delta^1(e)}}2 \tan\f{\phi_{\Delta^2(e)}}2}
\,.
\ee
This are exactly the geometric solutions to the loop polynomials $P_{T}$ that we derived earlier in  \eqref{GeometricSol} when considering the restriction of the pre-geometric parametrization (from complex edge lengths to real edge lengths satisfying the triangular inequalities). Here we have shown that they can be easily recovered as the saddle points of the generating function in the asymptotic regime of the 6j-symbol.

In fact, the key property of these saddle points is that they are scale-invariant. This is a generic property of this coherent spin network superpositions \cite{Bonzom:2012bn,Bonzom:2015ova}. More precisely, both the norm and phase of the saddle points $Y_{e}$ are expressed entirely as functions of the 2D and 3D angles of the tetrahedron, i.e. they are of the type $Y_{e} = F_{e}(j_1, \dotsc, j_6)$ where the $F_{e}$'s are invariant under homogeneous re-scalings of the spins $j_{e}$. This means that we do not simply have saddle points, but saddle lines. Each saddle line contributes an infinite redundancy in the series in the $j_{e}$'s, thus signals a pole in the generating function $\cZ_{T}$. This shows why  when the edge couplings $Y_{e}$ are such that they allow for saddle points according to the equation above  \eqref{saddlept}, then they are a pole of the spin network generating function $\cZ_{T}$, thus a zero for the 2D Ising model.

%
%
%


%
%
%
%

\section*{Conclusion}

We have applied the exact relation between the 2d Ising partition function and spin network evaluations \cite{Bonzom:2015ova,Westbury,Freidel:2012ji}, which expressed a duality between the 2d Ising model and 3d quantum gravity, to the simplest non-trivial case of the tetrahedral graph. In hat case, spin network evaluations on the tetrahedral graph are given by the 6j-symbols from the theory of recoupling of spins.

This led us to two original results. On the one hand, we obtained a self-duality identity satisfied by the generating function of 6j-symbols, which reflects the high/low temperature duality of the 2d Ising model. On the other hand, the asymptotics of the 6j-symbols for large spins, used in the quantum gravity context to describe the semi-classical regime of quantum geometry, leads to a geometric formula for the complex zeros of the 2d (inhomogeneous) Ising partition function on the tetrahedral graph. 

This provides an example for a more general method to derive the Fisher zeros for the 2d Ising model on arbitrary (3-valent) graph using the saddle point approximation for spin network evaluations (used, for instance, to probe the semi-classical limit of spinfoam path integral models for quantum gravity). The idea is to trade the zeros of the Ising partition function for the poles of the generating function for spin network evaluations. Indeed, considering a graph $\Gamma$, the generating function is (up to a numerical factor) the inverse squared of the Ising partition function:
\be
\cZ_{\Gamma}[\{Y_{e}\}_{e\in\Gamma}]
\propto
\f1{\cI_{\Gamma}[\{Y_{e}\}]^2}
\,,\qquad
\cI_{\Gamma}[\{Y_{e}\}]^2\propto\f1{\cZ_{\Gamma}[\{Y_{e}\}]}
\,,
\ee
where  the $Y_{e}$'s give the Ising couplings on the graph edges.
While the Ising partition function $\cI_{\Gamma}[\{Y_{e}\}]$ is a sum over spin up and spin down $\sigma_{v}=\pm$, the generating function $\cZ_{\Gamma}[\{Y_{e}\}]$ is defined by a series over half-integer spins $j_{e}\in\f\N2$. Studying the series by asymptotic methods, searching for saddle points, allows to identify the poles of $\cZ_{\Gamma}$ in the complex space and thus the zeros of $\cI_{\Gamma}$. We believe that this should lead to a general geometric formula for the Fisher zeros in terms of the embedding of 2d triangulations in the 3d space.
This illustrates a fruitful exchange of techniques between statistical physics and quantum gravity.

\section*{Acknowledgement}

EL is grateful to Prof. Kimyeong Lee for his hospitality at the Korean Institute for Advanced Study (KIAS) from March to May 2015,  during which most of the research work for this project has been done. VB was supported by the ANR MetAConc project ANR-15-CE40-0014.

\appendix
%

\section{Loop expansion of the 6j Generating function  and Racah formula}
\label{loopexpansion}

Let us start with the generating function for the 6j-symbols, which is equal to the inverse squared of the loop polynomial on the tetrahedron graph $T$, as we explain earlier in section \ref{sec:tetra}:
\be
\cZ_{T}[\{Y_e\}]
\,=\,
\sum_{\{j_{e\}}} \prod_{v}\Delta_v(j_e)\,
\sixj{j_{1}}{j_{2}}{j_{3}}{j_{4}}{j_{5}}{j_{6}} \,\prod_{e}Y_{e}^{2j_{e}}
=
\frac{1}{P_{T}[\{Y_e\}]^2}
\,,
\quad\textrm{with}\quad
P_{T}[\{Y_e\}]=1+\sum_{\substack{C\subset T \\ C\ne \varnothing}} \prod_{e\in C} Y_{e}
\,.
\nn
\ee
We define the cycle variables $L_C = \prod_{e\in C} Y_e$ as the product of the edge variables along the cycle $C$ in $T$. If we now expand the loop polynomial in terms of the cycle variables, the generating function admits an expansion onto loops,
\begin{equation}
\label{LCexpansion}
\frac{1}{P_{T}[Y_e]^2}
=
\sum_{\{M_C\}} (-1)^{\sum_C M_C}\,\big{(}\sum_C M_C + 1 \big{)}{ !} \,\prod_C \frac{L_C^{M_C}}{M_C!}
\end{equation}
where we sum over integers $M_C\in\N$, which are interpreted as the occupation numbers\footnotemark{} of the corresponding cycles.
\footnotetext{
This expansion can be understood as expressing the decomposition of arbitrary spins $j_{e}$ as tensor products of spins $\f12$, thus representing the 6j-symbols as superpositions of products of elementary loops carrying the minimal spin $\f12$.
}
Using the asymptotics $\ln M! \sim M\ln M + \mathcal{O}(M)$, we can approximate the summand as a function of the variables $M_{C}$ when those are large:
\be
\big{(}\sum_C M_C + 1 \big{)}{ !} \,\prod_C \frac{(-L_C)^{M_C}}{M_C!}
\underset{M_{C}\gg 1}{\sim}
e^{\Phi[\{M_{C}\}]}
\qquad
\textrm{with}
\quad
\Phi[\{M_{C}\}]
=
\sum_{C}M_{C}\ln \bigg{(}\f{-L_{C}\sum_{C'} M_{C'}}{M_{C}}\bigg{)}
\,,
\ee
and write the  saddle point equations $\pp_{M_{C}}\Phi=0$ to identify the stationary (and thus dominant) contributions to the series:
\begin{equation}
L_C = -\frac{M_C}{\sum_{C'} M_{C'}}
\,.
\end{equation}
As $P_{T}[Y_e] = 1 + \sum_C L_C$, this clearly gives all the roots of $P_{T}$ by analytic continuation to $M_C\in\mathbbm{C}$.
This is exactly the equation \eqref{LoopParametrization} which we derived earlier in section \ref{sec:PreGeometrical}, but we do not get the  constraint $M_{126} M_{135} M_{234} M_{456} = -\Bigl(\sum_C M_C\Bigr) M_{1245} M_{1346} M_{2356}$. This constraint reflects that we are working with seven cycle variables $L_{C}$ instead of the six edge variables $Y_{e}$, so the $L_{C}$ have to satisfy one constraint.

We can avoid adding this constraint by hand by performing the saddle point analysis directly in terms of the edge variables $Y_{e}$ instead of the $L_{C}$'s. In order to express the series in terms of the $Y_{e}$'s, we perform the following change of variables in the sum over $\{M_C\}$, analogous to \eqref{LengthsChangeofVar}:
\begin{equation}
j_e = \frac12 \sum_{C \ni e} M_C \quad \in \frac{\mathbbm{N}}{2}\qquad \text{and}\qquad n = \sum_C M_C.
\end{equation}
The reverse relations are easily found. If $t$ is a 3-cycle of $T$, we call $v(t)$ the vertex of $T$ opposite to $t$, then
\begin{equation}
M_t = n - J_{v(t)}\qquad \text{with} \qquad J_v = \sum_{e\ni v} j_e =\sum_{e\notin t}j_{e}
\,,
\end{equation}
where $J_{v}$  is the sum of the spins meeting at the vertex $v$, while if $\gamma$ is a 4-cycle:
\begin{equation}
M_\gamma = J_\gamma - n \qquad \text{with} \qquad J_\gamma = \sum_{e\in\gamma} j_e
\,.
\end{equation}
Those variables therefore are the equivalent of the variables $s_\gamma$ and semi-perimeters $s_{t^*}$ introduced in Section \ref{sec:PreGeometrical}, with the identification of twice the spins with the complex lengths $2j_e\leftrightarrow l_{e}$. 
The key property of this change of variables is that:
\begin{equation}
\prod_C L_C^{M_C} = \prod_e Y_e^{2j_e}
\,.
\end{equation}
This allows to translate the expansion \eqref{LCexpansion} in powers of $L_{C}$ into powers of $Y_{e}$:
\begin{equation}
\frac{1}{P(Y_e)^2} = \sum_{j_1, \dotsc, j_6} \biggl(\sum_n \frac{(-1) (n+1)!}{\prod_{v \in\Gamma} (n - s_{v})!\, \prod_{\gamma\in\Gamma} (s_\gamma - n)!} \biggr) \prod_e Y_e^{2j_e}
\,,
\end{equation}
where the sum over $n$ factorizes.
We have thus recovered Racah formula for the 6j-symbol,
\begin{equation}
\prod_{v}\Delta_v(j_e)\,
\sixj{j_{1}}{j_{2}}{j_{3}}{j_{4}}{j_{5}}{j_{6}} = \sum_n \frac{(-1) (n+1)!}{\prod_{v \in\Gamma} (n - s_{v})!\, \prod_{\gamma\in\Gamma} (s_\gamma - n)!}
\,.
\end{equation}
This shows that loop polynomial expression for the generating function of the 6j symbols is equivalent to Racah formula for the 6j-symbol, up to the change of variables described above.

Now, the saddle point equation for this sum over the $j_{e}$'s will provide the expected constraint on the $M_C$'s. This  reasoning explains why the constraint \eqref{SaddleEqRacah} is exactly the saddle point equation of Racah sum, as used in \cite{Gurau:2008yh}.


\bibliographystyle{bib-style}
\bibliography{spinfoam}

\end{document}